\documentclass[aps,prc,twocolumn,amsmath,floatfix,superscriptaddress]{revtex4-1}
\usepackage{lineno,isotope,mathrsfs}
\modulolinenumbers[5]
\usepackage{amsmath,bbold,amssymb,epsfig,bm,mathrsfs,feynmp}
\usepackage{color,slashed,nicefrac,exscale,multirow,soul}
\usepackage{times,txfonts,xcolor,graphicx,gensymb,tikz}
\usepackage[normalem]{ulem}
\usepackage[breaklinks,colorlinks,citecolor=blue]{hyperref}
\usepackage{ulem}

\begin{document}
\title{Constraining compact star properties with
nuclear saturation parameters}
\author{Jia Jie Li}
\email{jiajieli@itp.uni-frankfurt.de}
\affiliation{Institute for Theoretical Physics, J. W. Goethe University,
D-60438 Frankfurt am Main, Germany}
\author{Armen Sedrakian}
\email{sedrakian@fias.uni-frankfurt.de}
\affiliation{Frankfurt Institute for Advanced Studies, D-60438
Frankfurt am Main, Germany}
\affiliation{Institute of Theoretical Physics, University of Wroclaw,
50-204 Wroclaw, Poland
}
\begin{abstract}
A set of hadronic equations of state (EoSs) derived from relativistic
density functional theory and constrained by terrestrial experiments,
astrophysical observations, in particular by the GW170817 event,
and chiral effective field theory ($\chi$EFT) of neutron matter
is used to explore the sensitivity of the EoS parameterization
on the few nuclear matter characteristics defined at the saturation density.
We find that the gross properties of compact stars are most sensitive
to the isoscalar skewness coefficient $Q_{\text{sat}}$ and the
isovector slope coefficient $L_{\text{sym}}$ around saturation
density, since the higher order coefficients, such as $K_{\text{sym}}$,
are fixed by our model. More specifically, (i) among these
$Q_{\text{sat}}$ is the dominant parameter controlling both the
maximum mass and the radii of compact stars while $L_{\rm sym}$
is constrained somewhat by $\chi$EFT of neutron matter; (ii) massive
enough ($M\sim 2.0~M_{\odot})$ compact stars featuring both hyperons
and $\Delta$ resonances can be obtained if the value of $Q_{\text{sat}}$
is large enough; (iii) the emergence of $\Delta$'s reduces the
radius of a canonical mass ($M\sim 1.4~M_{\odot})$ compact star
thus easing the tension between the predictions of the relativistic
density functionals and the inferences from the X-ray observation
of nearby isolated neutron stars.
\end{abstract}
\date{\today}
\maketitle
%
%\linenumbers
%
%--------------------------------------------------------------------------
\section{Introduction}
\label{sec:intro}
%--------------------------------------------------------------------------
%
Compact stars are unique laboratories for studies of dense matter.
The hadronic core of a compact star extends from half up to a few times
the nuclear saturation density $\rho_{\text{sat}}$. Currently, the most
rigorous constraint on the high-density behavior of the equation of state
(EoS) comes from the observations of a few massive pulsars with masses
$\sim 2~M_{\odot}$~\cite{Demorest2010,Antoniadis2013,Fonseca2016}.
These observations set a lower bound on the maximum mass predicted by
any EoS of dense matter. The recent detection of gravitational waves
from the binary neutron star inspiral event GW170817~\cite{GW170817a}
allowed to place constraints on the tidal deformability of compact stars
and thus to put additional constraints on the EoS of dense matter~\cite{Annala2018,
Fattoyev2018,Most2018,Desoumi2018,GW170817b,Tews2018,Paschalidis2018,Lijj2019}.
The GW170817 event is complementary to the mass measurements indicated
above as it allows one to put constraints on the properties (specifically,
radius and deformability) of a canonical-mass ($M\sim 1.4~M_{\odot}$)
neutron star.

The details of the composition of compact stars at high densities are
not fully understood yet and the possibilities include hyperonization
~\cite{Glendenning1985,Glendenning1991,Schaffner1994,Huber1998,Vidana2001,
Weissenborn2012a,Weissenborn2012b,Massot2012,Colucci2013,Dalen2014,Oertel2015,
Tolos2016,Raduta2017}, the appearance of $\Delta$ resonances~\cite{Boguta1982,
Glendenning1985,Prakash1992,Schurhoff2010,Drago2014a,Drago2014b,Lijj2018b,
Lijj2019,Caibj2015,Zhuzy2016} and transition from hadronic to quark
matter~\cite{Haensel1986,Heiselberg1993,Weber2005,Bonanno2012,Alford2013,
Alford2017,Alvarez-Castillo2016,Alvarez-Castillo2017}; for recent reviews
see~\cite{Yagikent2017,Chatterjee2016,Sedrakian2017,Baym2018}.

Observational information on masses and radii does not resolve the
underlying composition of matter. In particular, the Bayesian
inferences~\cite{Steiner2010,Steiner2013,Lattimer2014,Ozel2016} of
these parameters from the data infer only the total pressure as a
function of density. However, many phenomena associated with neutron
stars, e.g., their cooling, depend in an essential way on the
composition of matter in the entire range from the crust to the core
of the star. Their modeling requires as an input microscopically
derived EoS or parametrizations thereof - a problem that we will
address in this work. The current observational programs focusing
on neutron stars combined with the nuclear physics modeling of their
interiors are aimed at unraveling the features of the matter compressed
to very high densities.

Among various possibilities mentioned just above, the hyperonization
of dense matter becomes a serious possibility because hyperons are
energetically favored as the density increases inside a neutron star;
for recent reviews see Refs.~\cite{Chatterjee2016,Yagikent2017,Baym2018}.
The onset of hyperons entails a considerable softening of the EoS and
thus reduces the maximum mass of corresponding sequences of compact
stars compared to those based on purely nucleonic EoS~\cite{Vidana2001,
Weissenborn2012a,Weissenborn2012b,Massot2012,Colucci2013,Dalen2014,
Oertel2015,Tolos2016,Lijj2018a}. The existence of new degrees of
freedom in the core of a neutron star cannot be confirmed or ruled
out so far on the basis of astrophysical observations alone.

The physics of nuclear systems at and somewhat below the saturation
density and zero temperature is well constrained by the studies of
finite nuclei. The EoS of isospin symmetrical nuclear matter around
saturation density is well constrained because physical observables
that are dominated by the isoscalar sector have been measured with
very high precision. On the other hand, the isovector sector remains
poorly determined as the measurements of the observables that are
sensitive to the isovector channel lack the necessary precision.
Pure neutron matter sets the limiting behavior of isovector properties
of nuclear matter. In particular, the neutron matter EoS, obtained by
solving the many-body Hamiltonian derived from chiral effective field
theory ($\chi$EFT), is expected to be reliable up to densities
$\sim 1.3\rho_{\text{sat}}$~\cite{Drischler2016}. This allows one to
gauge the phenomenological theories of isospin asymmetrical matter by
requiring that in the limit of pure neutron matter the {\it ab-initio}
results for the EoS are reproduced.

Clearly, any viable EoS must simultaneously satisfy the constraints
from experimental and theoretical studies of nuclear systems near the
saturation density and the observational constraints deduced from
studies of compact stars. In this work, we present a density functional
based parametrization of the dense matter EoS for the hadronic
matter that (i) reproduces the saturation properties of isospin-symmetric
nuclear matter; (ii) in the limit of pure neutron matter matches the
$\chi$EFT-based {\it ab initio} results for the EoS of neutron matter,
(iii) allows for strangeness in the form of hyperons as well as for
$\Delta$ resonances; (iv) produces compact star sequences with
$M_{\text{max}}\gtrsim 2~M_{\odot}$ and $R_{M_{1.4}} \lesssim 13.8$~km,
where $R_{M_{1.4}}$ is the radius of $1.4M_{\odot}$ mass star.
The key new feature of our study is the mapping of the density-functional
based EoS onto a generic one that is parametrized in terms of a few
observables of nuclear systems at saturation, which we call
{\it characteristic parameters or characteristics},
see Eq.~\eqref{eq:Taylor_expansion} below. Note that the low-order
characteristic parameters are known from the data on nuclei and are
often referred to as ``nuclear empirical parameters''. We use the
former term below to refer to the full parameter set entering this
equation, among which some are not constrained experimentally.

Similar explorations were previously carried out using Skyrme density
functionals in order to constrain the symmetry energy by evaluating
the neutron skin~\cite{Chenlw2010,Zhangz2013}, giant monopole resonances
~\cite{Chenlw2012}, and the electric dipole polarizability~\cite{Zhangz2014}.
Correlations between the critical density of $\Delta^-$ formation
and the maximum mass of compact stars within the nonlinear (NL) density
functional theory has also been studied in Refs.~\cite{Caibj2015,Caibj2017}.
Furthermore, the tidal polarizability of a neutron star has been applied
to constrain the symmetry energy within the NL-density functional
theory~\cite{Fattoyev2013}. Also, nucleonic EoS based on the Taylor
expansions around the saturation density has been applied to assess
the effect of the high-order characteristics~\cite{Margueron2018a,
Margueron2018b,Zhangnb2018a} as well as to put potential constraints
among them~\cite{Zhangnb2018a,Zhangnb2018b}.

The paper is organized as follows. In Sec.~\ref{sec:frame} we outline
the framework necessary to compute the stellar structure and the general
properties of asymmetric nuclear matter. In Sec.~\ref{sec:results} we
show how the uncertainties in the values of nuclear matter characteristics
at the saturation influence the parameters of the compact stars. This
is combined with the constraints on the available parameter space set
by the current theoretical and observational information. Finally,
Sec.~\ref{sec:summary} summarizes our concluding remarks.

%---------------------------------------------------------------------------
\section{Theoretical Framework}
\label{sec:frame}
%---------------------------------------------------------------------------

%---------------------------------------------------------------------------
\subsection{EoS of hadronic matter}
%---------------------------------------------------------------------------

We use here the standard form of the Hartree density functional in
which Dirac baryons are coupled via meson fields~\cite{Vretenar2005,
Mengjie2006}. The theory is Lorentz invariant and, therefore, preserves
causality when applied to high-density matter. The baryons interact
via exchanges of $\sigma, \omega$ and $\rho$ mesons, which comprise the
minimal set necessary for a quantitative description of nuclear
phenomena~\cite{Serot1997}. In addition, we consider two hidden-strangeness
mesons ($\sigma^\ast,\phi$) which describe interactions between
hyperons~\cite{Schaffner1994,Oertel2015,Lijj2018a,Raduta2017}.

The Lagrangian is given by the sum of the free baryonic and mesonic
Lagrangians~\cite{Lijj2018a,Lijj2018b}, which we do not write down,
and the interaction Lagrangian which reads
\begin{eqnarray}\label{eq:interaction_Lagrangian}
\mathscr{L}_{\text{int}}
&=& \sum_B \bar{\psi}_B\Big(-g_{\sigma B}\sigma-g_{\sigma^\ast B}\sigma^\ast
    -g_{\omega B}\gamma^\mu\omega_\mu-g_{\phi B}\gamma^\mu\phi_\mu \nonumber \\
&&  -g_{\rho B}\gamma^\mu\vec{\rho}_\mu\cdot\vec{\tau}_B\Big)\psi_B
    + \sum_D (\psi_B \rightarrow \psi^\nu_D),
\end{eqnarray}
where $\psi$ stands for the Dirac spinor and $\psi^\nu$ for the
Rarita-Schwinger spinor. Index $B$ labels the spin-1/2 baryonic
octet, which comprises nucleons $N\in\{n,p\}$, and hyperons
$Y\in\{\Lambda,\Xi^{0,-},\Sigma^{+,0,-}\}$, while index $D$ refers to the
spin-3/2 resonance quartet of $\Delta$'s ($\Delta\in\{\Delta^{++,+,0,-}\}$)
which are treated as Rarita-Schwinger particles~\cite{Pascalutsa2007}.
The mesons couple to the baryon octet and $\Delta$'s with the strengths
determined by the coupling constants $g_{mB}$ and $g_{mD}$, which are
functionals of the vector density. The Lagrangian~\eqref{eq:interaction_Lagrangian}
is minimal, as it does not contain (a) isovector-scalar $\delta$
meson~\cite{Rocamaza2011a} and (b) the $\pi$ meson and the tensor
couplings of vector meson to baryons (both of which arise in the
Hartree-Fock theories~\cite{Long2006,Long2007,Lijj2015,Lijj2016,
Lijj2018a}), which are beyond the scope of this paper.

In the nucleonic sector, the meson-nucleon ($mN$) couplings are given
by~\cite{Typel1999,Niksic2002}
\begin{align}
g_{m N}(\rho_v)=g_{m N}(\rho_{\text{sat}})f_{m N}(r),
\end{align}
where $r=\rho_v/\rho_{\text{sat}}$ and $\rho_v$ is the baryon vector
density. For the isoscalar channel, one has
\begin{align}\label{eq:isoscalar_coupling}
f_{m N}(r)=a_m\frac{1+b_m(r+d_m)^2}{1+c_m(r+d_m)^2}, \quad m = \sigma,\omega,
\end{align}
with $f_{m N}(1)=1$, $f^{\prime\prime}_{m N}(0)=0$ and
$f^{\prime\prime}_{\sigma N}(1)=f^{\prime\prime}_{\omega N}(1)$.
The density dependence for the isovector channel is taken in an
exponential form
\begin{align}\label{eq:isovector_coupling}
f_{mN}(r) = e^{-a_m (r-1)}, \quad m = \rho.
\end{align}
It is seen that if we fix in the Lagrangian~\eqref{eq:interaction_Lagrangian}
the baryon and meson masses to be (or close to) the ones in the vacuum
then properties of infinite nuclear matter can be computed uniquely in
terms of seven adjustable parameters. These are the three coupling
constants at saturation density ($g_{\sigma N}, g_{\omega N}, g_{\rho N}$),
and four parameters ($a_\sigma, b_\sigma, a_\omega, a_\rho$) that control
their density dependence.

The vector meson-hyperon ($mY$) couplings are given by the SU(6)
spin-flavor symmetric quark model~\cite{Oertel2015,Weissenborn2012a,
Colucci2013,Lijj2018a} whereas the scalar meson-hyperon couplings
are determined by fitting to certain preselected properties of
hypernuclear systems. We determine the coupling constants, $g_{\sigma Y}$,
using the following hyperon potentials in the symmetric nuclear matter
at saturation density $\rho_{\text{sat}}$~\cite{Feliciello2015,Gala2016}:
\begin{align}
\label{eq:hyp_potentials}
U^{(N)}_\Lambda = -U^{(N)}_\Sigma= -30~\text{MeV},\quad U^{(N)}_\Xi = -14~\text{MeV}.
\end{align}
Physically, the $\Lambda\Lambda$ bond energy provides a rough
estimate of the $U^{(\Lambda)}_\Lambda$ potential at the average
$\Lambda$ density $(\approx \rho_{\text{sat}}/5)$ inside a
hypernucleus~\cite{Vidana2001,Khan2015,Margueron2017}. We adopt
the value
\begin{align}
\label{eq:hyphyp_potentials}
U^{(\Lambda)}_\Lambda (\rho_{\text{sat}}/5) = - 0.67~\text{MeV},
\end{align}
which reproduces the most accurate experimental value to
date~\cite{Ahnjk2013}. This information we use to fix the
value of the coupling $g_{\sigma^\ast \Lambda}$. It
has been shown in Refs.~\cite{Khan2015,Margueron2017}
that the bond energy can be approximated by the
$U^{(\Lambda)}_\Lambda$ potential if the rearrangement
term in the mean field between double-$\Lambda$ and
single-$\Lambda$ hypernuclei is negligible. The coupling
of remaining hyperons $\Xi$ and $\Sigma$ to the $\sigma^*$
is constrained by the relation
$g_{\sigma^\ast Y}/g_{\phi Y} =
g_{\sigma^\ast\Lambda}/g_{\phi\Lambda}.$
Detailed discussions of hyperon potentials can be found,
e.g., in Refs.~\cite{Tolos2016,Fortin2017,Lijj2018a}.

The isoscalar meson-$\Delta$ ($m\Delta$) couplings are uncertain,
as no consensus has been reached yet on the magnitude of the
$\Delta$ potential in nuclear matter. The studies of the scattering
of electrons and pions off nuclei and photoabsorption which are
based on a phenomenological models~\cite{Alberico1994,Nakamura2010}
indicate that the $\Delta$ isoscalar potential $V_\Delta$ should
be in the range~\cite{Drago2014a}
\begin{equation}
\label{eq:scattering_constraint}
-30~\text{MeV} + V_N(\rho_{\text{sat}}) \lesssim V_\Delta(\rho_{\text{sat}})
\lesssim V_N(\rho_{\text{sat}}),
\end{equation}
where $V_N$ is the nucleon isoscalar potential. The studies of $\Delta$
production in heavy-ion collisions~\cite{Ferini2015,Cozma2016} suggest
a less attractive potential~\cite{Kolomeitsev2017},
\begin{equation}
\label{eq:HIC_constraint}
V_N(\rho_{\text{sat}}) \lesssim V_\Delta(\rho_{\text{sat}})
\lesssim 2/3V_N(\rho_{\text{sat}}).
\end{equation}
At the same time, the isovector meson-$\Delta$ couplings are
largely unknown. Below, we limit ourselves to the case where
$R_{\Delta\omega} = g_{\omega\Delta}/g_{\omega N} = 1.1$,
$g_{\rho\Delta}/ g_{\rho N} = 1.0$ and $g_{\sigma\Delta}$
is determined by fitting to the $\Delta$-potential at saturation
density $\rho_{\text{sat}}$. The value $R_{\Delta\omega} = 1.1$
allows one to obtain a physical solution for very attractive
$\Delta$-potentials~\cite{Lijj2018b}. Note that we assume
that the hyperon and $\Delta$ potentials scale with density as
the nucleonic potential, therefore their high-density behavior
is inferred from that of the nucleons. Such an assumption has
its justification in the quark substructure of these constituents
(where $\Delta$'s involve three-body bound states of light quarks
only, as nucleons, and strangeness-1 hyperons involve bound
states of two light and one heavy quark). However, first principle
computations which may confirm our assumption are still lacking.

Once the coupling constants are determined, one could compute the
EoS of the stellar matter by implementing the additional conditions
of weak equilibrium and change neutrality that prevail in neutron
stars. We further match smoothly our EoS for the core to that of
the crust EoS given in Refs.~\cite{Baym1971a,Baym1971b} at the
crust-core transition density $\rho_{\text{sat}}/2$. The integral
parameters of a compact star, in particular, the mass and the
radius, are then computed from the Tolman-Oppenheimer-Volkoff
(TOV) equations~\cite{Tolman1939,Oppenheimer1939}.

%---------------------------------------------------------------------------
\subsection{Characteristic parameters of nuclear matter}
%---------------------------------------------------------------------------
%
\begin{table}[tb]
\caption{The characteristic parameters of symmetric nuclear matter at
saturation density for DD-ME2 parametrization~\cite{Lalazissis2005}.
The bold parameters are those that can be calibrated by the density
functional alone. The definitions of parameters are as in Ref.~\cite{Margueron2018a}.
The $\rho_{\text{sat}}$ is in unit of fm$^{-3}$, $M^\ast_D$ in nucleon mass,
and the rest are in units of MeV.
}\setlength{\tabcolsep}{6.5pt}
\label{tab:SMP}
\begin{tabular}{cccccccccc}
\hline\hline
\multicolumn{8}{c}{Isoscalar characteristics} \\
\cline{4-8}$\rho_{\text{sat}}$ & $M^\ast_D$& &$E_{\text{sat}}$& &$K_{\text{sat}}$&$Q_{\text{sat}}$&$Z_{\text{sat}}$\\
\hline
     \textbf{0.152}&\textbf{0.57}& &$\textbf{-16.14}$& &\textbf{251.15} &\textbf{479}&4448 \\
\hline
\multicolumn{8}{c}{Isovector characteristics} \\
\cline{4-8}& & &$ E_{\text{sym}}$&$L_{\text{sym}}$&$K_{\text{sym}}$& $Q_{\text{sym}}$ & \\
\cline{4-8}
           & & &\textbf{32.31}&\textbf{51.27} &  $-87$  & 777 &  \\
\hline\hline
\end{tabular}
\end{table}

As is well known, the EoS of isospin asymmetric nuclear matter can be
expanded close to the saturation and the isospin symmetrical limit in
power series
\begin{align}
\label{eq:Taylor_expansion}
E(\chi, \delta) & \simeq  E_{\text{sat}} + \frac{1}{2!}K_{\text{sat}}\chi^2 + \frac{1}{3!}Q_{\text{sat}}\chi^3 \nonumber \\
                & + E_{\text{sym}}\delta^2 + L_{\text{sym}}\delta^2\chi
                  + {\mathcal O}(\chi^4,\chi^2\delta^2),
\end{align}
where $\chi=(\rho-\rho_{\text{sat}})/3\rho_{\text{sat}}$ and
$\delta = (\rho_{\text{n}}-\rho_\text{p})/\rho$. The coefficients
of the density-expansion given by the first line of
Eq.~\eqref{eq:Taylor_expansion} are known as the empirical
parameters of nuclear matter in the isoscalar channel, specifically,
the saturation energy $E_{\text{sat}}$, the incompressibility
$K_{\text{sat}}$, and the skewness $Q_{\text{sat}}$. The isovector
characteristics associated with the expansion away from the
symmetrical limit [the second line in Eq.~\eqref{eq:Taylor_expansion}]
are the symmetry energy parameter $E_{\text{sym}}$ and its slope
parameter $L_{\text{sym}}$. The higher-order terms in the expansion
\eqref{eq:Taylor_expansion}, which are not shown here, have been
studied in Refs.~\cite{Margueron2018a,Zhangnb2018a}.

It is then seen that, per definition, the various characteristics of
the bulk nuclear matter are the coefficients of the expansion of the
energy density close to the saturation density and isospin-symmetrical
limits (note that $\delta$ appears in even powers only). In order to
fully determine the parameters of our relativistic density functional,
we specify [in addition to the parameters appearing in
Eq.~\eqref{eq:Taylor_expansion}] the value of the Dirac mass $M^\ast_D$
at the saturation, which is important for a quantitative description
of finite nuclei, e.g., spin-orbit splitting.

Thus, given the five macroscopic characteristics in
Eq.~\eqref{eq:Taylor_expansion} together with the preassigned values
of $\rho_{\text{sat}}$ and $M^\ast_D$, we are in a position to
determine uniquely the seven adjustable parameters of the density
functional defined above. Having this in mind, our strategy would be
to vary individually these macroscopic characteristics within their
acceptable ranges and to examine the influence of these variations on
the EoS of dense matter and properties of compact stars. In this manner,
we explore the correlation(s) between specific properties of nuclear
matter and/or compact stars and each parameter entering
Eq.~\eqref{eq:Taylor_expansion}. Of particular interest are the quantities
which arise at a higher order of the expansion, specifically,
$Q_{\text{sat}}$ and $L_{\text{sym}}$. Their values are weakly
constrained by the conventional fitting protocol used in constructing
the density functionals, i.e., the procedure which involves usually
fits to nuclear masses, charge radii and neutron skins, see for
instance Refs.~\cite{Dutra2012,Sellahewa2014,Dutra2014,Margueron2018b}.
It is worthwhile to mention that there is a strong correlation
between $L_{\text{sym}}$ and the neutron skins~\cite{Horowitz2001,
Centelles2009,Rocamaza2011b,Chatterjee2017}. Unfortunately, the
uncertainties in the determination of neutron skins are large and
as a consequence, the experimental constraints on the theory are
weak~\cite{Abrahamyan2012,Tsang2012}.

For our analysis below we adopt as a reference the DD-ME2
parametrization~\cite{Lalazissis2005}. It has been tested on the
entire nuclear chart with great success and agrees with experimentally
known bounds on the empirical parameters of nuclear matter. In
Table~\ref{tab:SMP} we list the characteristic parameters of
symmetric nuclear matter at saturation density according to the
DD-ME2 parametrization.

The coefficients of the terms in the expansion
\eqref{eq:Taylor_expansion} that are higher than the second order
in $\chi$ and $\delta$ are highly model dependent~\cite{Dutra2012,
Sellahewa2014,Dutra2014,Margueron2018a}. For example, nonrelativistic
Skyrme/Gogny models predict negative $Q_{\text{sat}}$
value~\cite{Dutra2012,Sellahewa2014,Margueron2018a}, whereas
relativistic mean-field models predict positive $Q_{\text{sat}}$
value~\cite{Dutra2014,Margueron2018a}. Note that once the free
parameters of our density functional are fixed using the low-order
characteristics, these higher-order characteristics are {\it predicted}
by the density functional, i.e., these are not free parameters in the
present setup.

%--------------------------
\section{Results and discussions}
\label{sec:results}
%--------------------------
%
%---------------------------------------------------------------------------
\subsection{Low-density neutron matter}
%---------------------------------------------------------------------------
%
\begin{figure}[tb]
\centering
\ifpdf
\includegraphics[width = 0.45\textwidth]{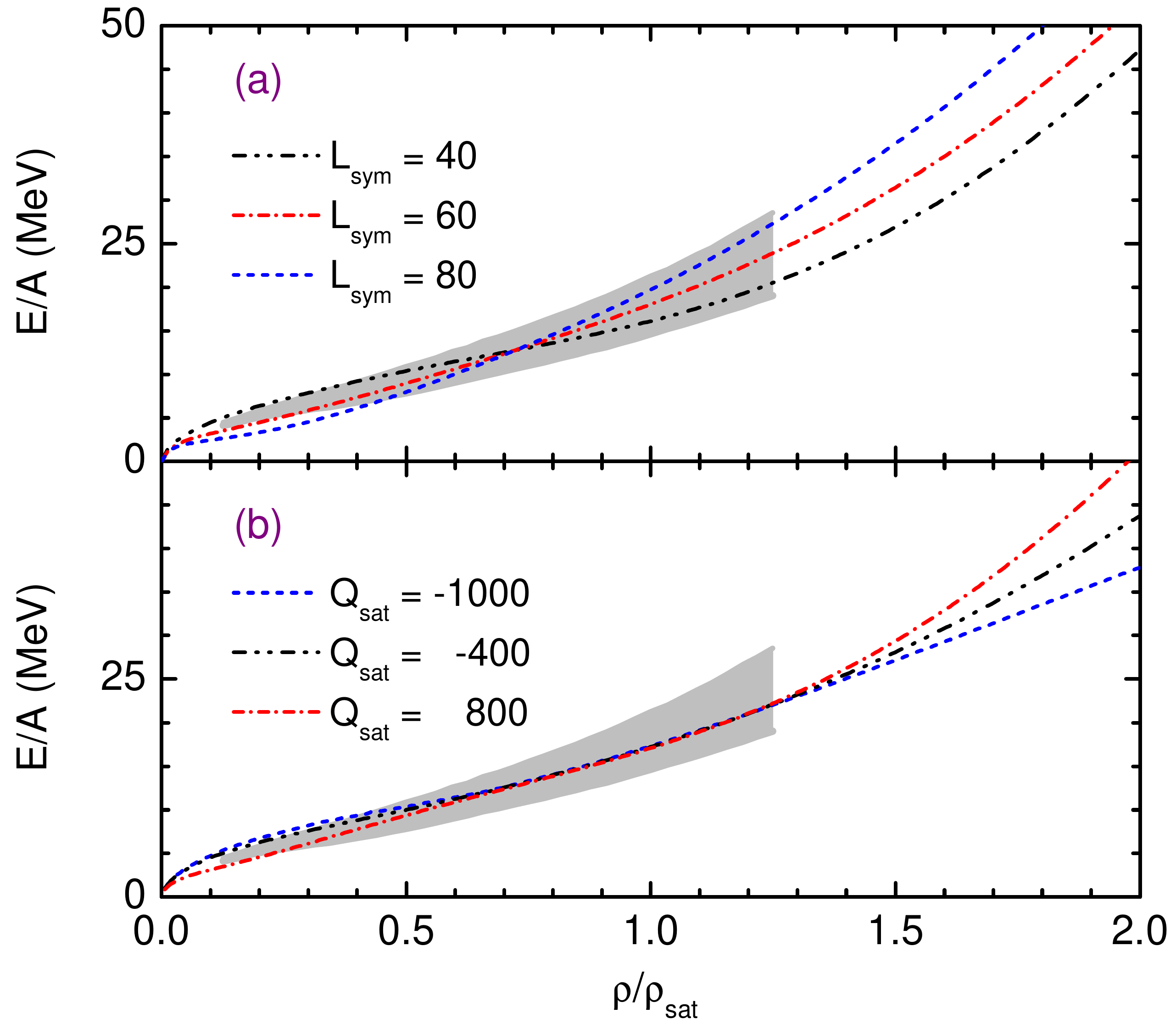}
\else
\includegraphics[width = 0.45\textwidth]{PNM.eps}
\fi
\caption{ Examples illustrating constraints on $L_{\text{sym}}$ and
$Q_{\text{sat}}$ set by $\chi$EFT. Shown are the energy per particle of
neutron matter as a function of $\rho/\rho_{\text{sat}}$ for different
values of $L_{\text{sym}}$~[MeV] (a) and $Q_{\text{sat}}$~[MeV] (b).
The shaded bands show the $\chi$EFT results~\cite{Drischler2016}.
In each panel, only the indicated parameter is varied, whereas the
remaining parameters are fixed at the default values.}
\label{fig:PNM}
\end{figure}

As outlined in the previous section, the nuclear matter EoS close to
the saturation can be characterized in terms of double-expansion
around the saturation density and isospin-symmetrical limit. The
coefficients of the expansion can be considered as {\it characteristic
parameters (or characteristics)} of nuclear matter, whereby the
parameters $E_{\text{sat}}$, $K_{\text{sat}}$, $E_{\text{sym}}$, and
$L_{\text{sym}}$, which are the coefficients of the dominant terms in
the expansion, have been studied extensively in the literature, for
reviews see~\cite{Liba2008,Rocamaza2018,Liba2018}. Correlations
between the input in the density functionals and these characteristics
have been established in a number of works~\cite{Vretenar2003,Chenlw2010,
Fattoyev2011,Rocamaza2013,Zhangz2014}. These studies suggest that
one can generate a set of density functional models by {\it varying
only one characteristic} while fixing the others. As mentioned above,
we consider as characteristics the five expansion coefficients in
Eq.~\eqref{eq:Taylor_expansion} plus the values of $\rho_{\text{sat}}$
and $M^\ast_D$ at saturation density. We then further restrict the
set of EoS by choosing only those which reproduce the result for
neutron matter at densities below and around saturation derived
from the {\it ab initio} calculations based on $\chi$EFT for
densities up to $\sim 1.3\rho_{\text{sat}}$~\cite{Drischler2016}.

While we fix the characteristics at saturation density, the nuclei
are most sensitive to the physics at densities that are below the
saturation density. Indeed, it has been recognized by several
authors~\cite{Furnstahl2002,Niksic2008,Trippa2008,Ducoin2011,
Zhangz2013,Colo2014} that a variety of nuclear models which
fit the properties of nuclear systems predict almost identical
values of symmetry energy for the density $\rho_c = 0.11$~fm$^{-3}$.
Motivated by this, we hold the value of the symmetry energy
$E_{\text{sym}}(\rho_c)$ [instead of $E_{\text{sym}}(\rho_{\text{sat}})$]
constant when $L_{\text{sym}}$ is being varied.

In Fig.~\ref{fig:PNM} we show the EoS which are compatible with the
neutron matter EoS and which lie within the allowed band region
obtained from studies based on $\chi$EFT. These EoS are obtained by
changing $L_{\text{sym}}$ (upper panel) or $Q_{\text{sat}}$ (lower
panel), while keeping all other characteristics at their default
values of DD-ME2 parametrization. It is seen that the
uncertainties in the values of these parameters allowed by the
$\chi$EFT have a minor influence on the behavior of the EoS at
subsaturation density. However, they significantly affect the
behavior of the EoS at higher densities (above $\sim 2\rho_{\text{sat}}$).
The energy of neutron matter below $\rho_c$ ($\rho_{\text{sat}}$)
becomes larger for the model with smaller $L_{\text{sym}}$
($Q_{\text{sat}}$)~\cite{Margueron2018a,Zhangnb2018a}.

We also illustrate in Fig.~\ref{fig:PNM} some typical cases for
the neutron matter EoS that are outside the $\chi$EFT band. For
$L_{\text{sym}} = 80$~MeV considerable deviation from $\chi$EFT
result in the very low-density regime is observed, although this
value is still consistent with the bounds $L_{\text{sym}}=58.7\pm28.1$~MeV
obtained from the combined analysis of astrophysical constraints and
terrestrial experiments~\citep{Oertel2017}. As seen in Fig.~\ref{fig:PNM},
the energy is slightly overestimated compared to the $\chi$EFT
calculations for $Q_{\text{sat}} = -1000$~MeV in the very low-density
regime. This shows that the influence of $Q_{\text{sat}}$ on the
behavior of the EoS at subsaturation density is vanishing for
$Q_{\text{sat}} \lesssim -500$~MeV. In the following, we shall restrict
our attention to those EoS models which satisfy the constraints on
low-density neutron matter from $\chi$EFT calculations~\cite{Drischler2016}.

The discussion above (see Fig.~\ref{fig:PNM}) is based on the
DD-ME2 parameterization. Since there is compensation between the
isoscalar and isovector channels, a change in the parameterization
for isoscalar channel will change the constraints for the isovector
channel, and vice versa. However, the change of the Lagrangian or
even the form of the functional (for example to the NL form~\cite{Fattoyev2013})
does not change the general features deduced above. In addition, our
setup does not allow us to vary freely the higher order parameters,
such as $K_{\text{sym}}$ or $Z_{\text{sat}}$ in Table~\ref{tab:SMP},
because once the low-order characteristic is fixed, the higher
order ones are the predictions of our density functionals. This
is in contrast to the models based entirely on Taylor
expansions~\cite{Margueron2018a,Margueron2018b,Zhangnb2018a,Zhangnb2018b},
where higher-order characteristics were varied at will.

%---------------------------------------------------------------------------
\subsection{Uncertainties in characteristics and compact stars}
%---------------------------------------------------------------------------
%
\begin{figure}[tb]
\centering
\ifpdf
\includegraphics[width = 0.48\textwidth]{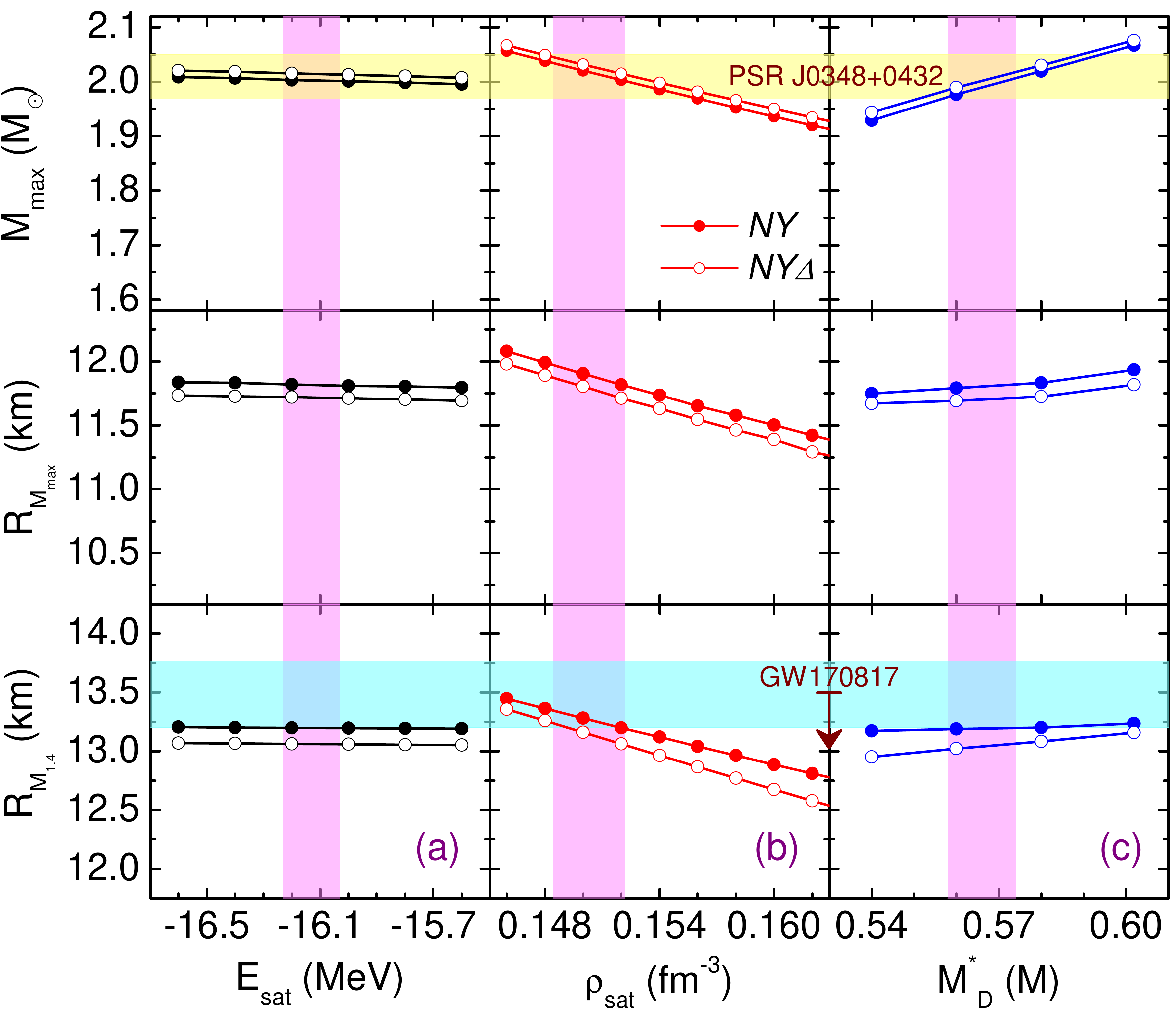}
\else
\includegraphics[width = 0.48\textwidth]{SUMa.eps}
\fi
\caption{Gross properties of compact stars for
nucleon--hyperon ($NY$) and nucleon--hyperon--Delta-resonance
($NY\Delta$) compositions. The maximum mass $M_{\text{max}}$ (upper
panels), the corresponding radii $R_{M_\text{max}}$ (middle panels),
and the radii $R_{M_{1.4}}$ for the canonical mass stars (lower
panels) are varied by tuning individually the energy $E_{\text{sat}}$
(a), the density $\rho_{\text{sat}}$ (b), and the Dirac mass $M^{\ast}_D$
(c), with the remaining parameters being fixed. The vertical shading
in each figure indicates the effect of varying the values of parameters
around their mean value considering 1$\sigma$ deviation. The yellow
shadings show the mass of PSR J0348+0432~\cite{Antoniadis2013}. The
light-blue shadings indicate the spreads of the upper limit on the
radius for a canonical 1.4$M_\odot$ mass neutron star set by recent
analysis of the tidal deformability determined from the GW170817
event~\cite{Annala2018,Fattoyev2018,Most2018,Desoumi2018,GW170817b,Tews2018}.}
\label{fig:SUMa}
\end{figure}
\begin{figure*}[tb]
\centering
\ifpdf
\includegraphics[width = 0.90\textwidth]{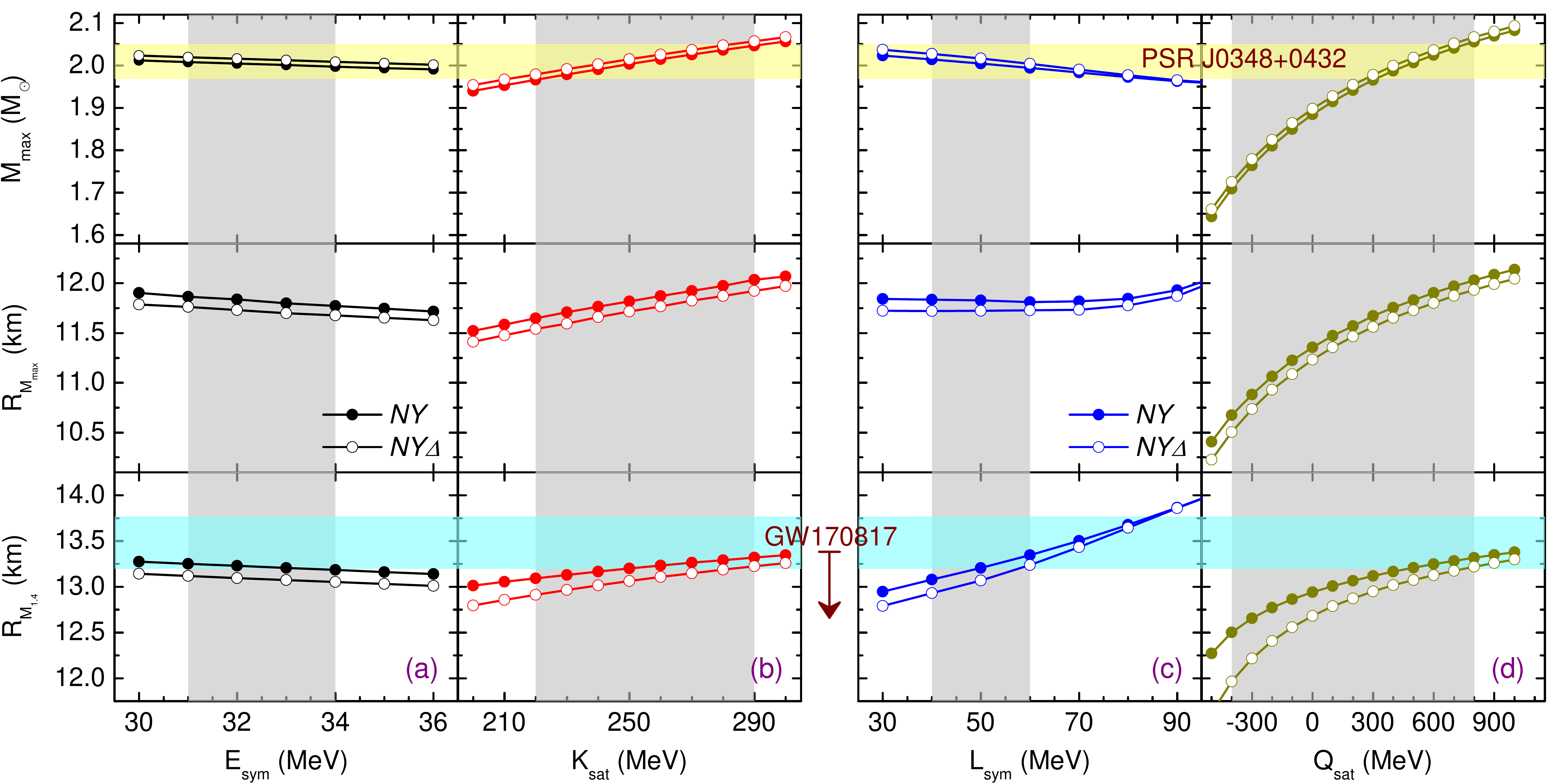}
\else
\includegraphics[width = 0.90\textwidth]{SUMb.eps}
\fi
\caption{Same as in Fig.~\ref{fig:SUMa}, but the panels show the
results for tuning of (a) symmetry energy $E_{\text{sym}}$, (b)
compression modulus $K_{\text{sat}}$ (c) slope of symmetry energy
$L_{\text{sym}}$ and (d) isoscalar skewness coefficient $Q_{\text{sat}}$.
The vertical shading in each figure indicates the constrains from
$\chi$EFT calculations.}
\label{fig:SUMb}
\end{figure*}

We now study the correlations between the gross properties of compact
stars and each nuclear characteristics at saturation density.
We base our exploration on the DD-ME2 interaction by varying individually
the seven characteristics within their empirical uncertainty ranges.
(Recall that we vary one characteristic at a time, i.e., all others
are held fixed at their default values defined by the DD-ME2
parametrization.) It is worthwhile to point out that although the five
macroscopic characteristics listed in Eq.~(\ref{eq:Taylor_expansion})
together with $\rho_{\text{sat}}$ and $M^\ast_D$ simultaneously affect
the EoS of dense matter, they are treated as independent of each other
in the present analysis.

Figures~\ref{fig:SUMa} and~\ref{fig:SUMb} show the gross properties
of compact stars with hyperon ($NY$) and Delta resonance ($NY\Delta$)
compositions. We vary individually the isoscalar characteristics
$E_{\text{sat}}$, $\rho_{\text{sat}}$, $M^\ast_D$, $K_{\text{sat}}$
and $Q_{\text{sat}}$, and the isovector characteristics $E_{\text{sym}}$
and $L_{\text{sym}}$. For illustrative purposes we fix the meson-$\Delta$
couplings by assuming the $\Delta$ potential satisfies the condition
$V_\Delta(\rho_{\text{sat}}) = V_N(\rho_{\text{sat}})$. It should be
mentioned that one has to modify all the 5 parameters in isoscalar
sector in order to vary the $\rho_{\text{sat}}$, $E_{\text{sat}}$,
and $M^\ast_D$, while one needs to modify only the 3 density-dependent
parameters instead to vary the characteristics $K_{\text{sat}}$ and
$Q_{\text{sat}}$. In this context, variations of $K_{\text{sat}}$ and
$Q_{\text{sat}}$ do not impact the meson-hyperon and meson-$\Delta$
couplings at nuclear saturation density.

If hyperons and no $\Delta$'s are admixed in the stellar matter, the
first hyperon to appear is the $\Lambda$, which is followed by the
$\Xi^-$ hyperon. The $\Sigma$ hyperons are disfavored due to their
repulsive potential at nuclear saturation density. This sequence of
hyperon thresholds is consistent with the recent relativistic
hypernuclear computations of Refs.~\cite{Weissenborn2012b,Fortin2017,
Lijj2018a}. As a result, the hyperons appear in compact stars with
masses with $M_{\text{max}} \gtrsim 1.5M_{\odot}$, i.e., masses larger
than the canonical pulsar mass. When $\Delta$ resonances are taken into
account by taking $V_\Delta(\rho_{\text{sat}}) = V_N(\rho_{\text{sat}})$,
$\Delta^-$ is the first isobar to be populated around 2$\rho_{\text{sat}}$;
its number density grows and reaches the number density of protons at
$\sim 3 \rho_{\text{sat}}$. At even higher densities it is gradually
replaced by the $\Xi^-$ hyperons around 4$\rho_{\text{sat}}$. It has
been shown that $\Delta$ resonances soften the EoS at low densities but
stiffen it at high densities~\cite{Lijj2018b}. (The corresponding
particle content of matter will be discussed below in Fig.~\ref{fig:Frac1}.)
It is thus seen that the overall trends are rather similar when varying
individually the characteristics for $NY$ and $NY\Delta$ matter.
The difference between the two compositions is clearly reflected
in the radius of a canonical neutron star. Note that in the entire
parameter space considered, the purely nucleonic EoS models always
predict a maximum mass of neutron star which is larger than $2M_{\odot}$.

It is seen from Fig.~\ref{fig:SUMa} that the maximum mass of a star
$M_{\text{max}}$ (the corresponding radius $R_{M_{\text{max}}}$)
decreases with $E_{\text{sat}}$ and $\rho_{\text{sat}}$, while it
increases with $M^\ast_D$. The radius for a canonical star $R_{M_{1.4}}$
exhibits similar correlation. These features indicate that the gross
properties of compact stars are sensitive to the values of $\rho_{\text{sat}}$
and $M^\ast_D$. Since the parametrizations presented in Fig.~\ref{fig:SUMa}
all satisfy the $\chi$EFT constraint for neutron matter, we show instead
the 1$\sigma$ deviations that are evaluated from the available
density-dependent relativistic mean-field (DD-RMF)
parametrizations (DD-ME~\cite{Niksic2002,Lalazissis2005},
DD~\cite{Typel2005,Klahn2006,Typel2010}, PKDD~\cite{Long2004} and
TW99~\cite{Typel1999}). It is clearly seen that this model is well
constrained with respect to the characteristics $\rho_{\text{sat}}$,
$E_{\text{sat}}$, and $M^\ast_D$ within $\sim 2\%$. Therefore,
the effect of varying the value of $\rho_{\text{sat}}$ (or $M^\ast_D$)
around the mean value at the level of 1$\sigma$ deviation is not significant.
It is worthwhile to mention that even the lowest order characteristics,
for instance, the $E_{\text{sat}}$ and $\rho_{\text{sat}}$ could be
different among different type of models, and the differences could be
larger than the standard deviations. Indeed, while nonrelativistic
models predict $\rho_{\text{sat}} \simeq 0.160 \pm 0.004$~fm$^{-3}$~\cite{Dutra2012,Sellahewa2014},
the relativistic models have a significantly smaller value
$\rho_{\text{sat}} \simeq 0.150 \pm 0.003$~ fm$^{-3}$~\cite{Dutra2014,Lijj2018a}. In this
context, the difference in the saturation density from relativistic
and nonrelativistic models yields already considerable effects on
gross properties of compact stars; see Fig.~\ref{fig:SUMa}(b).

We now turn our attention to the assessment of the effects of higher
order characteristics which are shown in Fig.~\ref{fig:SUMb}. The
vertical shading indicates the constraints from $\chi$EFT calculations.
Besides this, we have checked that all the constrained parameter sets
can reasonably reproduce the binding energies and charge radii of a
number of (semi-)closed-shell nuclei with $\sim 2\%$ relative deviation.
As seen from Fig.~\ref{fig:SUMb} (a) and (c), the maximum mass $M_{\text{max}}$ is
independent of the symmetry energy $E_{\text{sym}}$, while it shows a
weak negative correlation with the slope parameter $L_{\text{sym}}$.
The corresponding radius of the maximum-mass star $R_{M_{\text{max}}}$
is essentially independent on the value of $L_{\text{sym}}$ (and $E_{\text{sym}}$),
while the radius of a canonical neutron star $R_{M_{1.4}}$ is strongly
and almost linearly dependent on the value of $L_{\text{sym}}$. It is
interesting to note that the key two astrophysical constraints available
presently, i.e., $M_{\text{max}} \gtrsim 2.0M_\odot$ and
$R_{M_{1.4}} \lesssim 13.8$~km, favor small $L_{\text{sym}}$. As pointed
out in Refs.~\cite{Lattimer2001,Horowitz2001,Fortin2016}, there is a
correlation between the radius of 1.4$M_{\odot}$ stars and the magnitude of
$L_{\text{sym}}$. However, once the constraints placed by $\chi$EFT
calculations are implemented, the isovector characteristics
$E_{\text{sym}}$ and $L_{\text{sym}}$ have a very small influence on
the gross properties of compact stars. For example, the variations in
the maximum mass turn out to be of the order of $0.05~M_\odot$. The
variations in the radius of a canonical neutron star are of the order
of 0.4~km.

We further find that $M_{\text{max}}$, $R_{M_{\text{max}}}$ and
$R_{M_{1.4}}$ display positive correlation with the isoscalar
characteristics $K_{\text{sat}}$ and $Q_{\text{sat}}$, as shown
in Fig.~\ref{fig:SUMb} (b) and (d), respectively. The correlations
are almost linear for $K_{\text{sat}}$ but more complex for
$Q_{\text{sat}}$. While isoscalar skewness $Q_{\text{sat}}$
induces variation in both the maximum mass and radius, it largely
controls the maximum mass of compact stars because it is most
effective in modifying the EoS at supra-saturation densities. The
seemingly stronger correlation between $M_{\text{max}}$ and
$Q_{\text{sat}}$ compared to the correlations of $M_{\text{max}}$
with the other six variables is because of the relatively larger
uncertainty in $Q_{\text{sat}}$. In fact, the quality of the resultant
model depends not only on the form of the functional but, in addition,
on the data used for its calibration. Indeed, even within the same
functional form, the spread in values of $Q_{\text{sat}}$ is very
large, covering the range $\sim-500 < Q_{\text{sat}} < 500$~MeV
~\cite{Typel1999,Lalazissis2005}. The constraint,
$M_{\text{max}} \gtrsim 2.0M_\odot$, favors larger values for
$Q_{\text{sat}}$ (or $K_{\text{sat}}$), but the constraint,
$R_{M_{1.4}} \lesssim 13.8$~km, favors smaller values for
$Q_{\text{sat}}$ (or $K_{\text{sat}}$). Notice also the significant
reduction of $R_{M_{1.4}}$ for negative values of $Q_{\text{sat}}$
as seen in Fig.~\ref{fig:SUMb} (d).
We present a set of alternative parametrizations that preserve
these values of $\rho_{\text{sat}}$, $E_{\text{sat}}$, and $M^\ast_D$
as DD-ME2, but produce different values of $K_{\text{sat}}$ and/or
$Q_{\text{sat}}$ in Table~\ref{tab:Varying_KQ} of the Appendix.

It is interesting to notice that the maximum mass gradually saturates
at the value~$2.1M_\odot$ with an increase of the $Q_{\text{sat}}$.
We conclude that our study indicates an upper limit $\sim 2.1M_\odot$
on the maximum mass of compact stars with hyperon mixing. This prediction
will be further confirmed below, specifically in Fig.~\ref{fig:CQL}.
Interestingly, the value we find is compatible with the most recent
inferences on the maximum mass of neutron stars,
$M_{\text{max}} \sim 2.17M_\odot$~\cite{Margalit2017,Shibata2017,
Rezzolla2018}.

\begin{figure}[tb]
\centering
\ifpdf
\includegraphics[width = 0.48\textwidth]{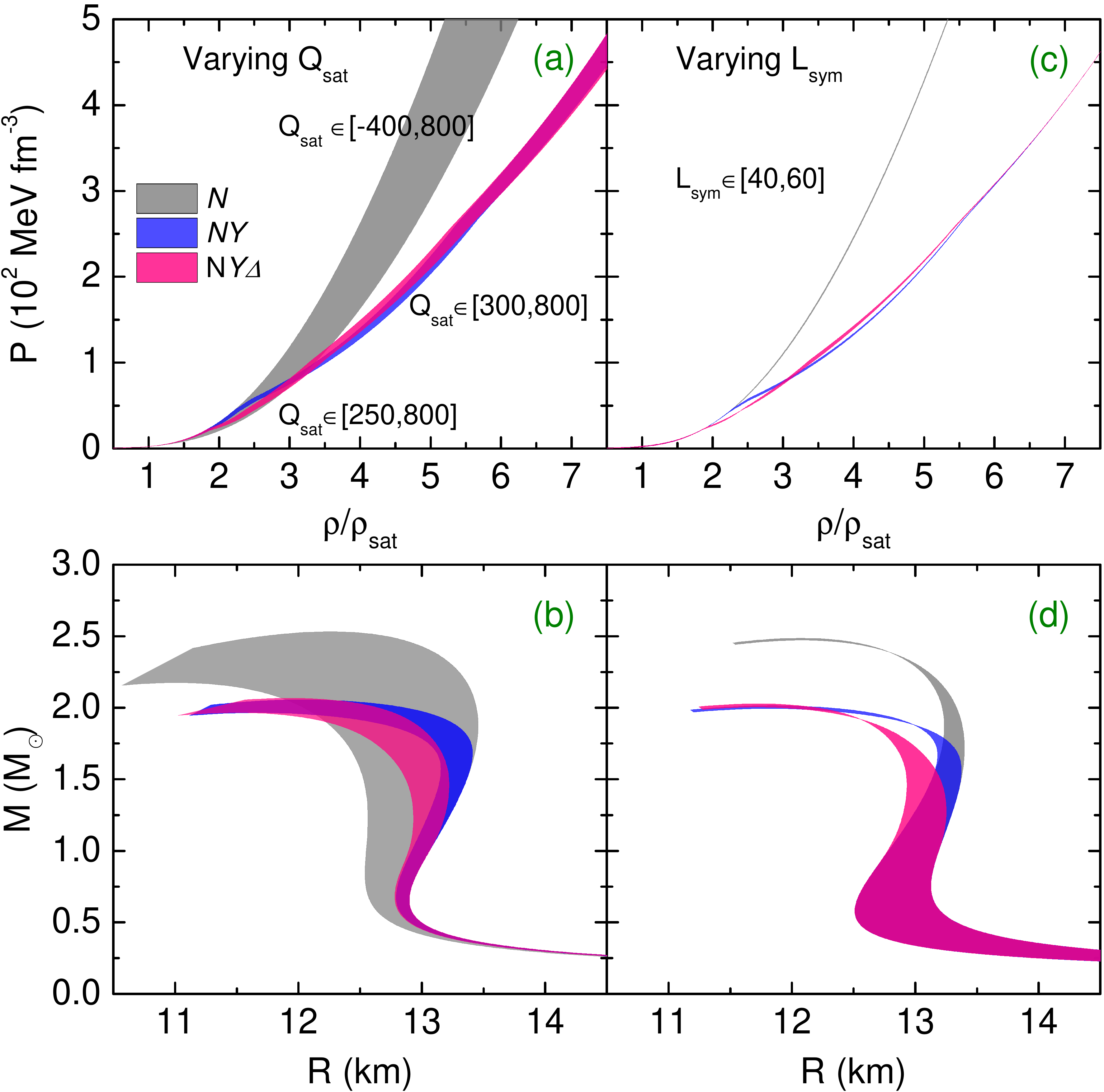}
\else
\includegraphics[width = 0.48\textwidth]{EOS1.eps}
\fi
\caption{ EoS models and MR relations for $N$, $NY$, and $NY\Delta$
compositions of stellar matter. The bands are generated by varying
the parameters $Q_{\text{sat}}$~[MeV] (a, b) and $L_{\text{sym}}$~[MeV]
(c, d). The ranges of  $Q_{\text{sat}}$ and $L_{\text{sym}}$ allowed
by $\chi$EFT and maximum mass constraints are indicated in the figures.}
\label{fig:EOS1}
\end{figure}

We now compare purely nucleonic compact stars with those which contain
hyperonic matter. To support a purely nucleonic star with a mass of
about 2.0$M_\odot$, $Q_{\text{sat}}$ needs to be just
$Q_{\text{sat}} \gtrsim -650$~MeV, leading to a value of $R_{M_{1.4}}$
can be small as $\sim 11.8$~km. Once one allows for hyperonization,
$Q_{\text{sat}}$ has to be, at least, as large as $\backsim 300$~MeV.
Thus, once the value of the maximum mass of a compact star is pinned down,
it will put a stringent upper limit on the $Q_{\text{sat}}$ parameter.
However, such a limit will largely depend on the composition of matter.

\begin{figure}[tb]
\centering
\ifpdf
\includegraphics[width = 0.48\textwidth]{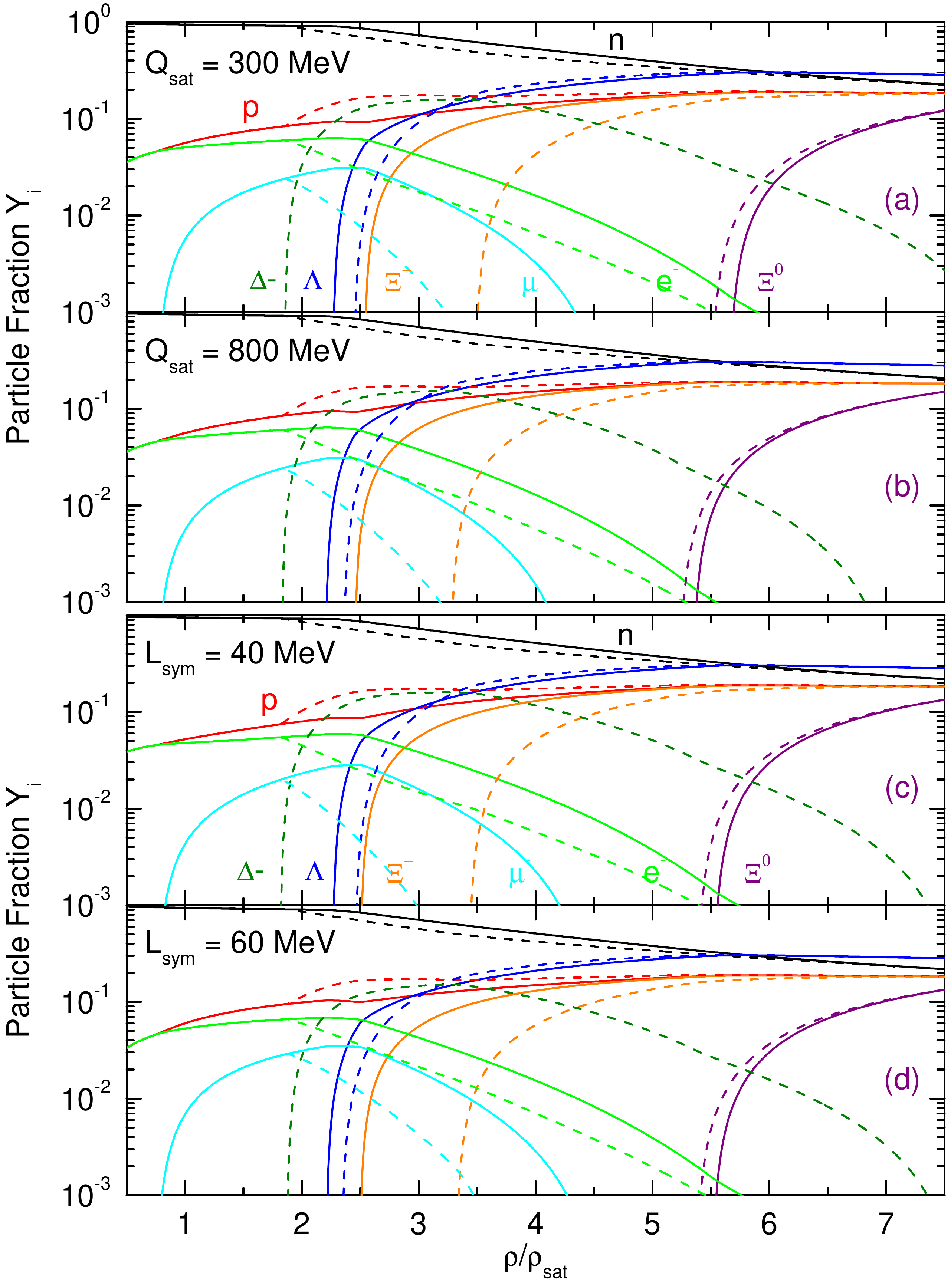}
\else
\includegraphics[width = 0.48\textwidth]{Frac1.eps}
\fi
\caption{Effects of varying the values of $Q_{\text{sat}}$ [MeV] (a, b)
  and $L_{\text{sym}}$ [MeV] (c, d) on particle fraction in
  $NY$ (solid lines)  $NY\Delta$ (dashed lines) stellar matter.}
\label{fig:Frac1}
\end{figure}

In Fig.~\ref{fig:EOS1} we show the EoS models of stellar matter and the
mass-radius (hereafter MR) relations for purely nucleonic, hyperon
admixed and hyperon-$\Delta$ admixed matter for the allowed ranges of
$Q_{\text{sat}}$ (left panels) and $L_{\text{sym}}$ (right panels).
It is clearly seen that the same microscopical and astrophysical
constraints lead to different EoS and MR relation depending on the
assumed particle content. The appearance of hyperons and/or
$\Delta$ resonances softens the EoS from baryon density
$\rho \sim 2.5\rho_{\text{sat}}$, which corresponds to the threshold
of the $\Lambda$ and /or $\Delta$ production. Furthermore, the
displayed MR relations show that $L_{\text{sym}}$ has an appreciable
effect on the radii of less massive stars ($M \lesssim 1.4M_\odot$),
whereas the $Q_{\text{sat}}$ has more significant effect on the radii
of heaver stars ($M \gtrsim 1.4M_\odot$). The canonical neutron stars
are just at the intersection where the effects of $Q_{\text{sat}}$ and
$L_{\text{sym}}$ on the radii are comparable, which implies that some
combinations of $Q_{\text{sat}}$ and $L_{\text{sym}}$ can lead to the
same $R_{M_{1.4}}$. Therefore, $Q_{\text{sat}}$ or $L_{\text{sym}}$
values alone are insufficient to characterize the low-density (up to
around $2\rho_{\text{sat}}$) behavior of EoS within relativistic
density functional theory. Our conclusion is consistent with that
in recent metamodeling for the nucleonic EoS~\cite{Margueron2018b}.
Notice however that in our models the hyperons/resonances could
appear already in canonical neutron stars.

Finally, we examine the effect of varying the value of
$Q_{\text{sat}}$ (upper panel) and $L_{\text{sym}}$ (lower panel) on
particle fraction which is shown in Fig.~\ref{fig:Frac1}. By changing
the value of $Q_{\text{sat}}$ in the interval [300,800] MeV which
corresponds to $\sim 50\%$ variations around its default value $\sim~480$~MeV
from DD-ME2, we observe that the effect of changing $Q_{\text{sat}}$
on the onset density of $\Delta^-$ and $\Lambda$ appears to be rather
small, while its effect on the onset density of $\Xi^{-,0}$ is more
visible. This is because $Q_{\text{sat}}$ characterizes the
medium- and high-density behavior of the isoscalar component of EoS.
As a result, the particle fractions shown in Fig.~\ref{fig:Frac1} (a)
and (b) differ to some for $\rho \gtrsim 3.5\rho_{\text{sat}}$.
Varying the value of $L_{\text{sym}}$ in the interval [40,60] MeV
which corresponds to $\sim 20\%$ variations around its default
DD-ME2 value $\sim 50$~MeV, we find that a larger value of $L_{\text{sym}}$
pushes up the threshold density of $\Delta^-$, while the onsets of
$\Lambda$ and $\Xi^-$ are shifted down.
Since within our model the isospin asymmetry of stellar matter
generally decreases with the increase of density, and the coupling
constant of vector meson decreases exponentially with density
[see Eq.(\ref{eq:isovector_coupling})], the isovector field is
largely suppressed at a higher density. As a result, the particle
fractions shown in Fig.~\ref{fig:Frac1} (c) and (d) become identical
for density $\gtrsim 4\rho_{\text{sat}}$.

In closing this section let us note that the value of $M_{\text{max}}$
for a compact star is basically determined by the isoscalar
characteristics of the EoS, i.e., $\rho_{\text{sat}}$,
$K_{\text{sat}}$, and $Q_{\text{sat}}$. The so-called ``hyperon
puzzle''~\cite{Chatterjee2016} is therefore mainly related to the
isoscalar skewness coefficient $Q_{\text{sat}}$ that characterizes the
medium- and high-density behavior of EoS. On the other hand, the
radius of the star is determined by both the isoscalar and isovector
characteristics of the EoS.  The constraints on neutron matter EoS
coming from $\chi$EFT do not allow for a wide variation of
$L_{\text{sym}}$, therefore one is left with the potential variations
of $Q_{\text{sat}}$ for the determination of the radius of a
1.4$M_{\odot}$ star. Thus, we conclude that the observations of
massive compact stars and advanced determinations of stellar radii
will potentially constrain the value of $Q_{\text{sat}}$ within
our model setup. It should be stressed again that the theoretically
inferred value of $Q_{\text{sat}}$ depends on the composition of
matter, and to a certain extent, the detailed form of the density
functional.

%---------------------------------------------------------------------------
\subsection{Observational constraints in the $Q_{\text{sat}}$-$L_{\text{sym}}$ plane}
%---------------------------------------------------------------------------
%
\begin{figure}[tb]
\centering
\ifpdf
\includegraphics[width = 0.48\textwidth]{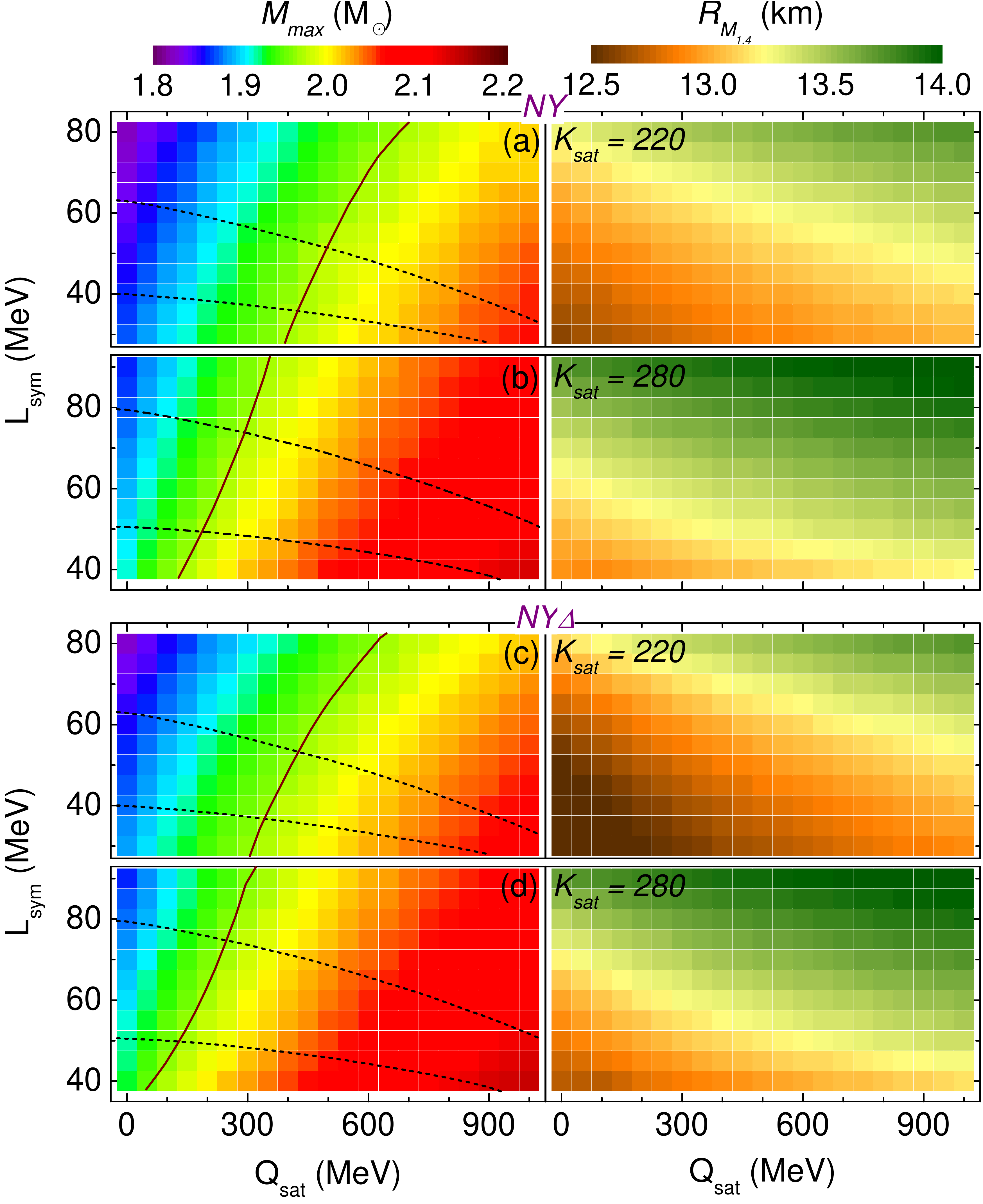}
\else
\includegraphics[width = 0.48\textwidth]{CQL.eps}
\fi
\caption{Contour plots for the gross properties of compact stars in
  the parameter space spanned by $L_{\text{sym}}$ and $Q_{\text{sat}}$
  (both in MeV) with two fixed values of $K_{\text{sat}} = 220$ and
  280~MeV. Shown are the maximum mass and the radius of a canonical
  1.4$M_{\odot}$ mass star with $NY$ (a, b) and $NY\Delta$ (c, d)
  compositions. The solid lines indicate the configurations that have
  a maximum mass equal to 1.97$M_{\odot}$. The dashed lines show the
  constrains from $\chi$EFT calculations. }
\label{fig:CQL}
\end{figure}

Having established some general trends by varying only one of the
parameters (i.e. only one of the dimensions in the parameter space) we
would like to explore next the parameter space when two dimensions are
varied. Our focus will be on the characteristics $Q_{\text{sat}}$ and
$L_{\text{sym}}$ which are less constrained so far. We use alternative
parametrizations that produce $K_{\text{sat}} = 220$ and 280~MeV,
respectively, but preserve these values of $\rho_{\text{sat}}$,
$E_{\text{sat}}$, and $M^\ast_D$ as DD-ME2. Figure~\ref{fig:CQL} shows
the value of the maximum-mass star, and the radius of a canonical
1.4$M_{\odot}$ star with $NY$ and $NY\Delta$ compositions, computed
by simultaneously varying both $Q_{\text{sat}}$ and $L_{\text{sym}}$.

Comparing Fig.~\ref{fig:CQL} (a) and (b), we observed that, (i)~to
satisfy the constraints imposed by $\chi$EFT, the value of
$L_{\text{sym}}$ has to be smaller for the EoS model which has small
$K_{\text{sat}}$, in order to enhance the contribution from the
symmetry energy at very low density, see Fig.~\ref{fig:PNM} (a);
(ii)~to support compact stars with the maximum mass
$M_{\text{max}} \gtrsim 1.97M_{\odot}$, the value of $Q_{\text{sat}}$
has to be larger for the EoS model which has small $K_{\text{sat}}$,
to compensate the smaller contribution from the $K_{\text{sat}}$;
(iii)~the maximum masses $M_{\text{max}}$ (radii of 1.4$M_{\odot}$
stars) predicted by EoS models with $K_{\text{sat}} = 220$~MeV are
typically $\sim 0.1M_{\odot}$ ($\sim0.3$~km) smaller than those by
EoS models with $K_{\text{sat}} = 280$~MeV; as already observed in
Fig.~\ref{fig:SUMb} (b). As a consequence, the uncertainties in
the isovector characteristics will impact the uncertainty intervals
of the isoscalar characteristics, and vice versa; the uncertainty
in a lower-order characteristic will impact somewhat the uncertainty
interval of a higher-order characteristics. Such an interplay
between the characteristic parameters have been discussed previously
in the metamodeling approach to nuclear EoS~\cite{Margueron2019}.

\begin{figure}[tb]
\centering
\ifpdf
\includegraphics[width = 0.48\textwidth]{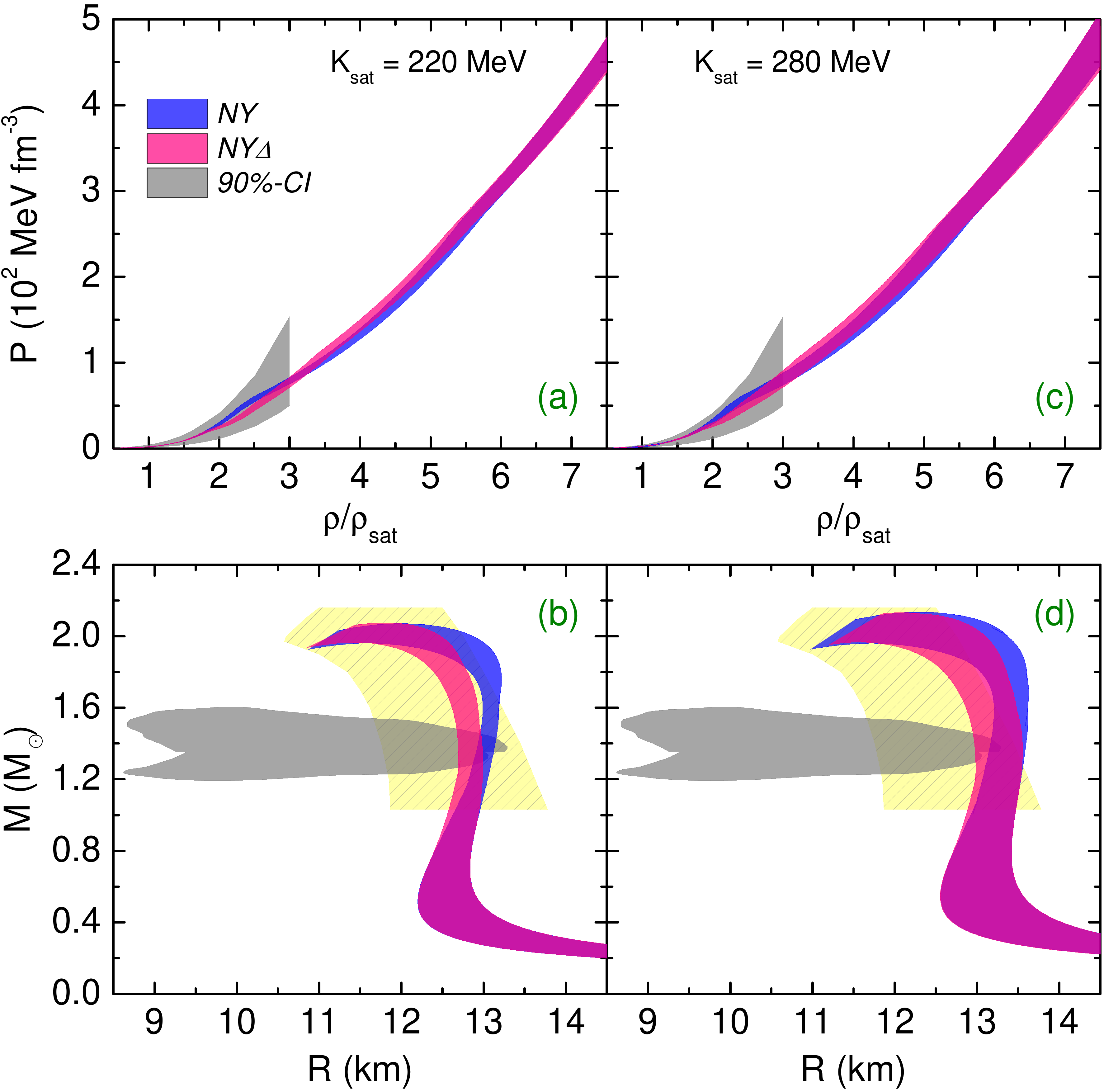}
\else
\includegraphics[width = 0.48\textwidth]{EOS2.eps}
\fi
\caption{EoS models and the corresponding MR relations for $NY$ and
  $NY\Delta$ compositions and for $K_{\text{sat}}=220$~MeV (a,b) and
  $K_{\text{sat}}=280$~MeV (c, d) within the allowed parameter space
  spanned by $Q_{\text{sat}}$ and $L_{\text{sym}}$ (both in MeV). In
  the top panels, gray shading represents the 90\% posterior credible
  level (90\%CI) estimated from the binary neutron star merger event
  GW170817~\cite{GW170817b}. In the bottom panels, gray shading
  represents the posterior for the mass and radius of each binary
  component using EoS-insensitive relations~\cite{GW170817b}, while
  the yellow shading indicates the $2\sigma$ region of radii
  inferred in the analysis of Ref.~\cite{Most2018}.}
\label{fig:EOS2}
\end{figure}

Consider now the radius of a canonical 1.4$M_{\odot}$ star with $NY$
composition. For those EoS models that satisfy the $\chi$EFT and
$M_{\text{max}} \gtrsim 2M_{\odot}$ constraints, the predicted radius
of a 1.4$M_{\odot}$ star spans from 12.8~km (defined by the EoS with
$K_{\text{sat}} = 220$, $L_{\text{sym}} \approx 35$, and
$Q_{\text{sat}} \approx 450$~MeV) to 13.6~km (defined by the EoS with
$K_{\text{sat}} = 280$, $L_{\text{sym}} \approx 60$, and
$Q_{\text{sat}} \approx 800$~MeV).  Notice that the values
$K_{\text{sat}} = 220$~MeV and $L_{\text{sym}} \approx 35$~MeV adopted
here are already very close to the current lower bound of constraints
on them placed by the combined analysis of terrestrial
experiments~\cite{Vretenar2003,Oertel2017}. Therefore, the
parameter space left for further reduction of the radius of a
1.4$M_{\odot}$ star appears to be rather small.

When the $\Delta$ resonances are taken into account and one sets
$V_{\Delta} (\rho_{\text{sat}}) = V_N (\rho_{\text{sat}})$ the
magnitude of $Q_{\text{sat}}$ may be taken to be slightly smaller
($\sim 100$~MeV), than in the case of $NY$ matter in order to
obtain a 2$M_{\odot}$ star. Indeed the inclusion of $\Delta$
resonances results in a larger maximum mass~\cite{Lijj2018b}.

\begin{figure}[tb]
\centering
\ifpdf
\includegraphics[width = 0.48\textwidth]{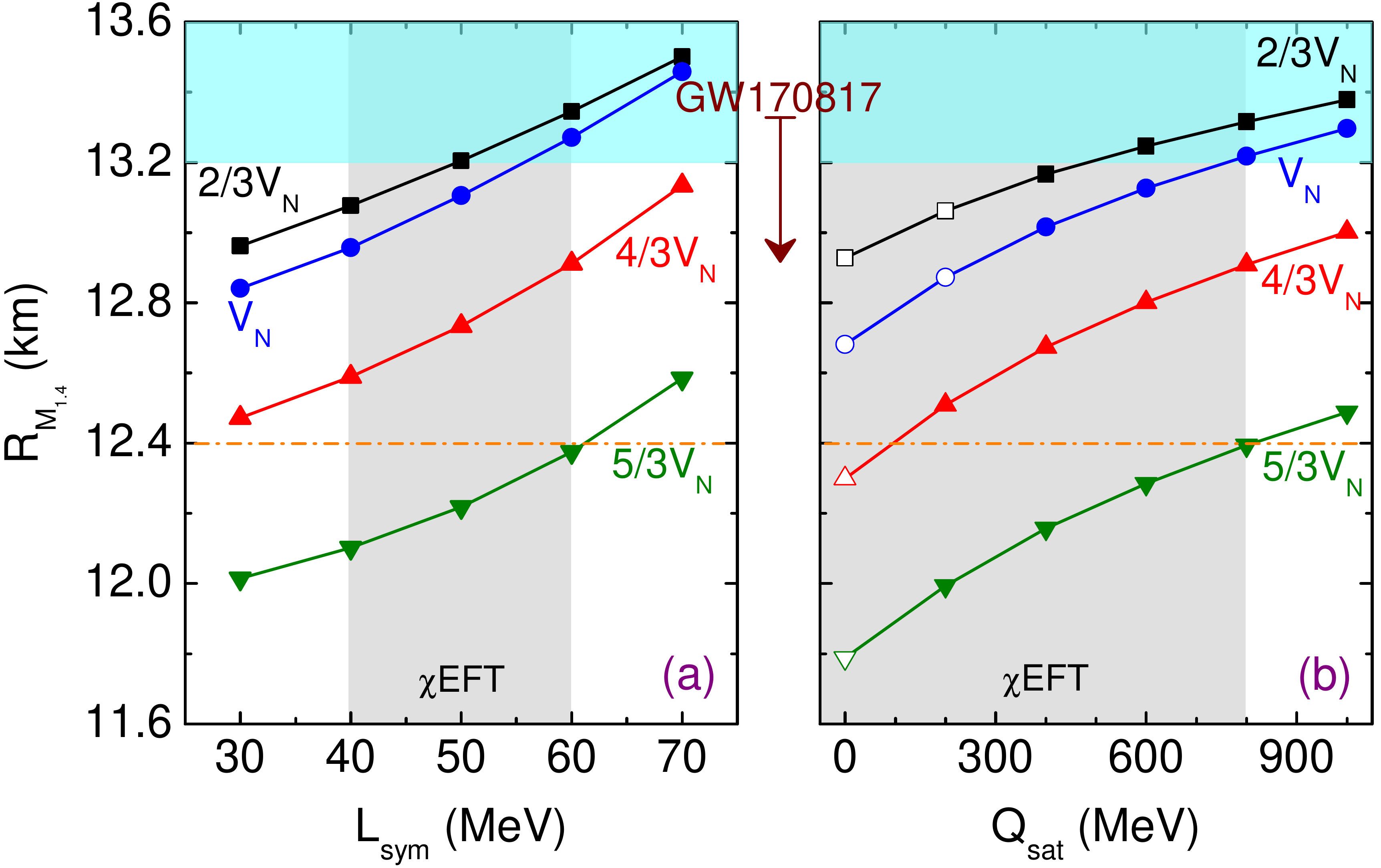}
\else
\includegraphics[width = 0.48\textwidth]{CRD.eps}
\fi
\caption{Radii of canonical mass stars $R_{M_{1.4}}$ versus (a) the
  slope of symmetry energy $L_{\text{sym}}$ and (b) the isoscalar
  skewness coefficient $Q_{\text{sat}}$ at saturation density. The
  open symbols mark the cases where the maximum-mass is below
  1.97$M_\odot$. The vertical shadings indicate the constrains from
  $\chi$EFT calculations~\cite{Drischler2016}. The light-blue shadings
  show the spreads of the upper limit for canonical 1.4$M_\odot$
  neutron stars set by recent analysis of the tidal deformability
  determined by GW170817 event~\cite{Annala2018,Fattoyev2018,Most2018,
  Desoumi2018,GW170817b,Tews2018}. The dashed lines mark the most
  likely value of 12.4~km set in Ref.~\cite{Most2018}.}
\label{fig:CRD}
\end{figure}

In Fig.~\ref{fig:EOS2} we summarize the EoS models (upper panels) and
the corresponding MR relations (lower panels) for $NY$ and $NY\Delta$
compositions with $K_{\text{sat}}=220$~MeV (left panels) and
$K_{\text{sat}}=280$~MeV (right panels) respectively, restricting the
($Q_{\text{sat}}$, $L_{\text{sym}}$) space to that shown in
Fig.~\ref{fig:CQL}.  We also show the constrains on the EoS obtained
at the 90\% posterior credible level (90\%CI) from the binary neutron
star merger event GW170817~\cite{GW170817b}, the posterior for the
mass and radius of each binary component using EoS-insensitive
relations in Ref.~\cite{GW170817b}, as well as the probable ($2\sigma$
region) radii of neutron stars estimated from a very large range of
hadronic EoS by imposing constraints on the maximum mass and the tidal
deformability~\cite{Most2018}. Since the exotic degrees of freedom
were not considered in Ref.~\cite{GW170817b}, the band corresponding
to 90\%CI constraints on the high-density regime are not shown
here. As can be seen from Fig.~\ref{fig:EOS2} (a) and (c), for the
low-density region $0.5\leqslant \rho/\rho_{\text{sat}} \leqslant 3$,
our collection of EoS are fully consistent with the inference of
Ref.~\cite{GW170817b}. The radii predicted by those models for a star
with the canonical mass 1.4$M_\odot$ lie close to the upper range of
the radii inferred from the analysis of tidal deformability from the
binary neutron star inspiral event GW170817~\cite{GW170817b,Most2018}.
The MR relations generated by EoS models which have
$K_{\text{sat}} = 220$~MeV appear to be in better agreement with the
$2\sigma$ domain inferred in Ref.~\cite{Most2018}.
However, it should be noted that, below the onset density of
hyperons/resonances ($\sim2.5\rho_{\text{sat}}$), our EoS models span a
small region of the inferred band, namely our EoS models do not
represent all possible models compatible with GW170817, but only a
subset of them.

Finally, it is worthwhile to note that if the vector meson-hyperon
couplings are drawn from within the SU(3) flavor symmetric
model~\cite{Weissenborn2012b,Lijj2018a}, rather than SU(6) spin-flavor
symmetric model, then one can obtain stiffer EoS at high
densities (and less hyperon-rich matter) which would increase the
maximum mass by up to 0.3$M_\odot$. In this case, the value of
$Q_{\text{sat}}$ can be reduced by several hundred without violating
the astrophysical mass constraint.

%---------------------------------------------------------------------------
\subsection{Canonical mass stars with small radii}
%---------------------------------------------------------------------------
%
\begin{figure}[tb]
\centering
\ifpdf
\includegraphics[width = 0.48\textwidth]{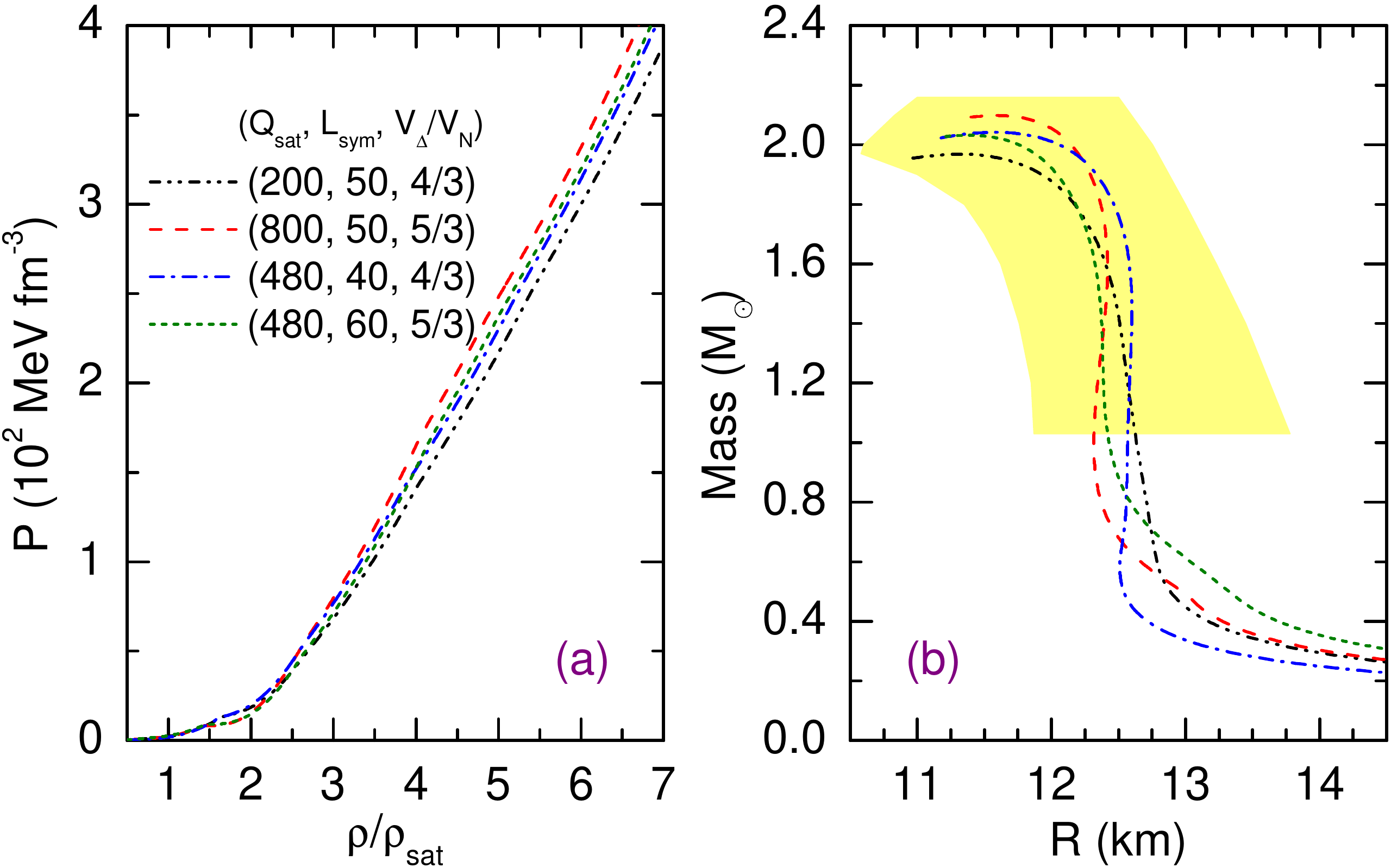}
\else
\includegraphics[width = 0.48\textwidth]{EOS3.eps}
\fi
\caption{ EoS models and the corresponding MR relations for $NY\Delta$
  stellar matter in the parameter space ($Q_{\text{sat}}$,
  $L_{\text{sym}}$, $V_\Delta$) (in MeV) that support a 1.4$M_\odot$
  neutron star with radius about 12.5 km. The yellow shading
  represents the probable ($2\sigma$ region) radii of neutron
  stars estimated in Ref.~\cite{Most2018}.}
\label{fig:EOS3}
\end{figure}

As mentioned in the previous section, the constraints on the tidal
deformability have allowed for the determination of the statistically
most probable radius of a 1.4$M_\odot$ neutron stars. For instance, by
imposing constraints on the maximum mass and on the dimensionless
tidal deformability, it has been shown that a purely hadronic neutron
star has $12.0 \leqslant R_{M_{1.4}} \leqslant 13.5$~km with a
2$\sigma$ confidence level, with a most likely value of
12.4~km~\cite{Most2018}. In Ref.~\cite{Desoumi2018} the binary neutron
star mergers with different prior choices of masses have been analyzed.
Using Bayesian parameter estimation the authors concluded that the
radius range is $8.9\leqslant R_{M_{1.4}} \leqslant 13.2$~km, with an
average value of $\bar{R}_{M_{1.4}}=10.8$~km~\cite{Desoumi2018}.

The possibility that hyperonic stars have small radii in the range
above is as exciting as it is challenging for nuclear theory.
Notice that small radii demand a sufficiently soft EoS below
2-3$\rho_{\text{sat}}$, while the observed large masses require that
the same EoS must be able to evolve into a stiff EoS at high
densities. In Ref.~\cite{Lijj2018b}, it was augured that $\Delta$
resonances soften the EoS at low densities but stiffen it at high
densities, resulting in significantly reduced radii and larger maximum
masses of compact stars.  The $\Delta$ resonances are therefore an
interesting degree of freedom for the modeling of small-radius
stars. The effects of $\Delta$ resonances have been illustrated for the
case of $V_{\Delta} = V_{N}$ in previous sections. We next explore
further their effects by varying the $\Delta$-potential.

Figure~\ref{fig:CRD} shows the radius for a 1.4$M_{\odot}$ star as a
function of $L_{\text{sym}}$ ($Q_{\text{sat}}$) and $V_{\Delta}$,
while the remaining characteristic parameters are set as default
values in Table~\ref{tab:SMP}. As expected, the changes in the
$\Delta$-potential $V_{\Delta}$ have a stronger effect on the radius.
The appearance of $\Delta$ resonances reduces the radius of a canonical
star by up to 1~km for a reasonably attractive $\Delta$-potential
$V_{\Delta}=5/3V_{N}$, thus producing a radius which is closer to
the inferred most likely value 12.4~km obtained by Ref.~\cite{Most2018}.
It is worth noticing that the reduction is not sensitive to the values
of $Q_{\text{sat}}$ and/or $L_{\text{sym}}$.

\begin{figure}[tb]
\centering
\ifpdf
\includegraphics[width = 0.48\textwidth]{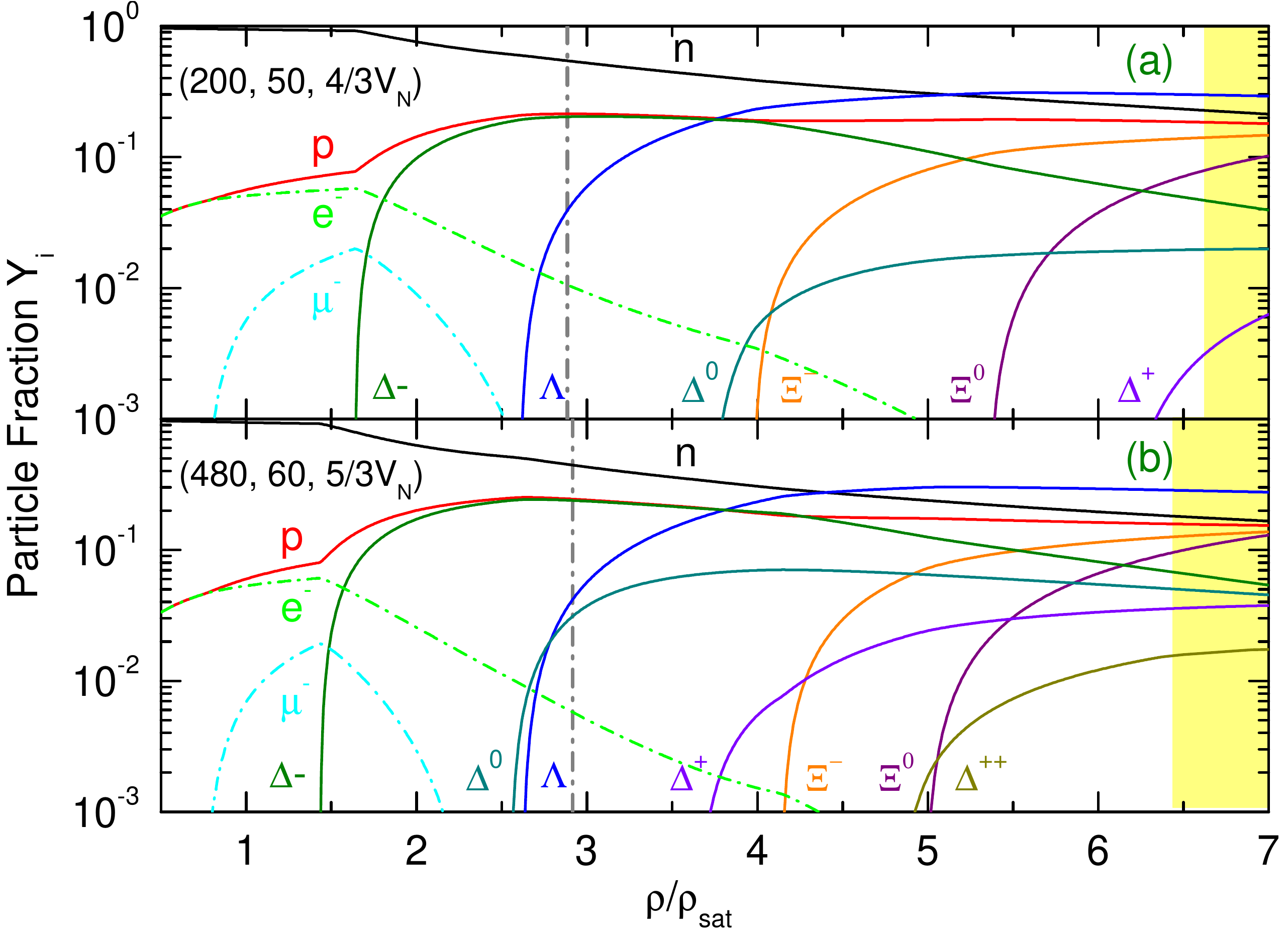}
\else
\includegraphics[width = 0.48\textwidth]{Frac2.eps}
\fi
\caption{ Particle fraction for $NY\Delta$ stellar matter in the parameter space
  ($Q_{\text{sat}}$, $L_{\text{sym}}$, $V_\Delta$) (in MeV) that support a
  1.4$M_\odot$ neutron star with radius about 12.5 km. The thick vertical lines
  indicate the central density of the respective canonical 1.4$M_\odot$ neutron
  star, the yellow shadings show densities beyond the maximum mass configurations.}
\label{fig:Frac2}
\end{figure}

We present in Fig.~\ref{fig:EOS3} several EoS models and the corresponding
MR relations for $NY\Delta$ matter in the parameter space ($Q_{\text{sat}}$,
$L_{\text{sym}}$, $V_\Delta$) that reproduce a 1.4$M_\odot$ neutron
star with a radius about 12.5~km. The particle fractions for two EoS
models are shown in Fig.~\ref{fig:Frac2} for illustration. As can be
seen in such configuration the $\Delta^-$ appears already at
$\sim1.5\rho_{\text{sat}}$, and $\Delta^0$ appears at
intermediate densities. At the very central part of a canonical neutron
star, the concentration of $\Delta^-$ resonance is close to that of protons.

\begin{figure}[tb]
\centering
\ifpdf
\includegraphics[width = 0.42\textwidth]{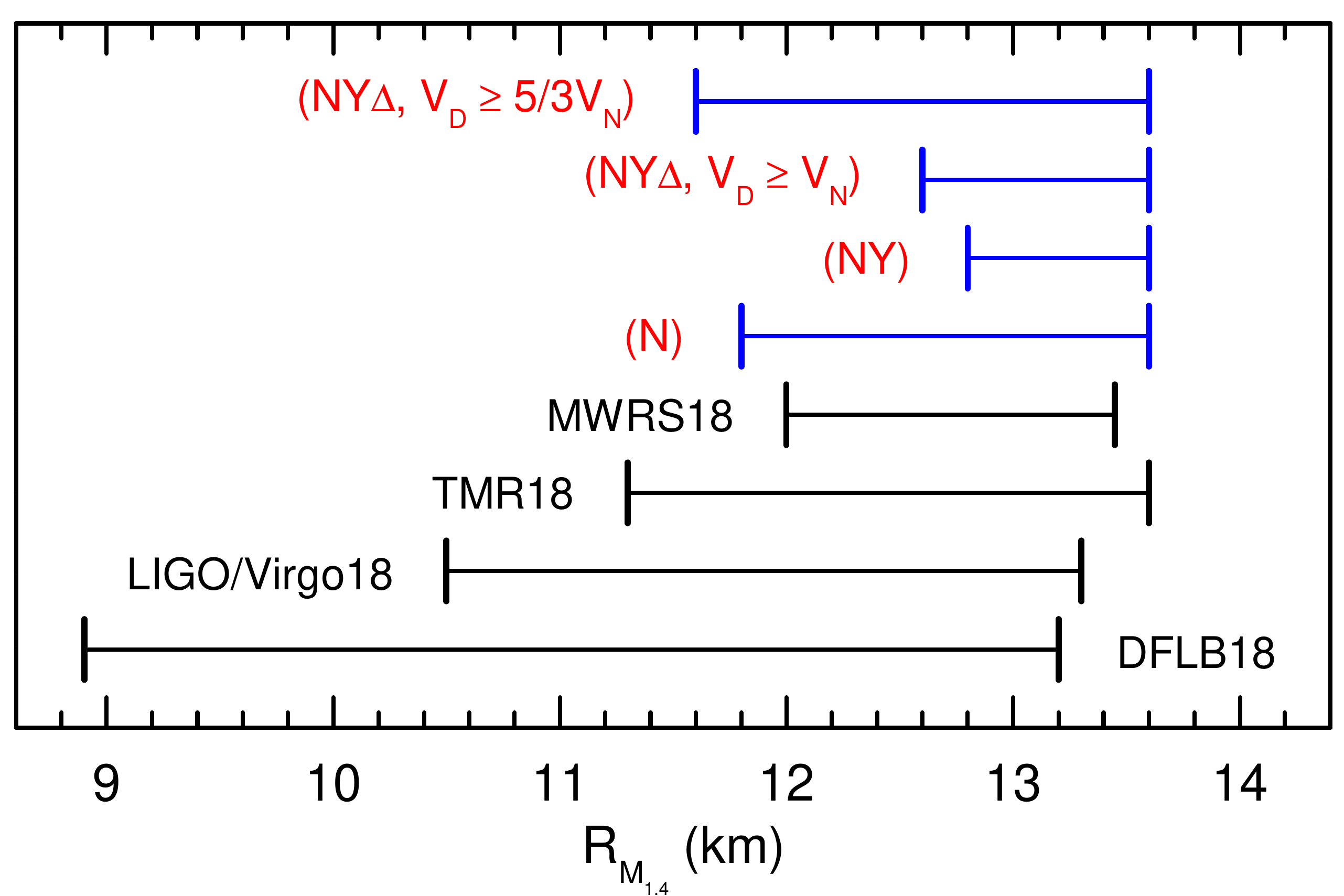}
\else
\includegraphics[width = 0.42\textwidth]{CR.eps}
\fi
\caption{Constraints on the radii of canonical neutron stars from
the analysis of the GW170817 event using hadronic EoS models
(MWRS18~\cite{Most2018}, DFLB18~\cite{Desoumi2018}, LIGO/Virgo18~\cite{GW170817b} 
and TMR18~\cite{Tews2018}), and from present models
assuming purely nucleonic ($N$), hyperon admixed ($NY$), and
hyperon-$\Delta$ admixed ($NY\Delta$) particle composition for the stellar
matter. The limits are set on the parameter space spanned by $(K_{\text{sat}},
Q_{\text{sat}}, L_{\text{sym}})$ restricting $K_{\text{sat}}$ within the
range $ [220, 280]$~MeV. The value of the $\Delta$-potential is taken either
$V_{\Delta} \geq V_{N}$ or $V_{\Delta} \geq 5/3V_{N}$.}
\label{fig:CR}
\end{figure}

Finally, in Fig.~\ref{fig:CR} we show the limits on the radius of a
1.4$M_\odot$ canonical neutron star from the work which used the data
on GW170817 event, along with the radii obtained from our EoS models
assuming purely nucleonic ($N$), hyperon admixed ($NY$), and
hyperon-$\Delta$ admixed ($NY\Delta$) particle composition.  We recall
that our limits are set on the
$(K_{\text{sat}}, Q_{\text{sat}}, L_{\text{sym}})$ parameter space, by
restricting $K_{\text{sat}}\in [220, 280]$~MeV. The value of the
$\Delta$-potential $V_{\Delta}$ is varied from $2/3V_{N}$ to
$5/3V_{N}$. Further reduction of the radius up to 2~km can be obtained
if one further decrease the $V_{\Delta}$, see Ref.~\cite{Lijj2018b}
for a detailed discussion. It is clearly seen from Fig.~\ref{fig:CR}
that our estimate of the upper limit of $R_{M_{1.4}}$ is consistent
with other analyses~\cite{Most2018,Desoumi2018,GW170817b,Tews2018}.
The upper limit is in fact set by the purely nucleonic EoS, whereas
the lower limit essentially depends on the assumed particle composition.
Our hyperon-$\Delta$ admixed EoS models ($V_{\Delta} \geqslant 5/3V_{N}$)
place a lower limit of about 11.6~km. Interestingly, this limit is
rather close to the one set by the purely nucleonic EoS models.
This underlines our argument that while the EoS uniquely
determines the MR relation, it does not allow one to extract
information on the composition of dense matter. A canonical-mass
star with small radius could be interpreted not only as a purely
nucleonic object, but also as hypernuclear star admixed with $\Delta$
resonances~\cite{Lijj2018b,Lijj2019} or, alternatively, a hybrid star
containing a quark matter core~\cite{Paschalidis2018,Alvarez2019}.
Furthermore, it appears that $R_{M_{1.4}} \lesssim 11$~km is marginally
compatible with our present knowledge of the nuclear, hypernuclear and
$\Delta$ resonance physics data. Canonical mass stars with small radii
(less than 11~km) may therefore indicate the possibility of hadron-quark
phase transition at density around $2\rho_{\text{sat}}$~\cite{Annala2018,
Paschalidis2018,Most2018,Tews2018,Alford2017}.

%-------------------------------------------------------------------
\begin{table*}[tb]
\centering
\caption{Meson masses and meson-nucleon coupling constants in the
  DD-ME2 parametriztation~\cite{Lalazissis2005}, whereby $g_{mN}$
  refer to the values at the saturation density.}
\setlength{\tabcolsep}{12pt}
\label{tab:DDME2}
\centering
\begin{tabular}{ccccccccc}
\hline\hline
$m_\sigma$ & $m_\omega$  & $m_\rho$  &$g_{\sigma N}$&$g_{\omega N}$ &$g_{\rho N}$&            & $a_\rho$  \\
 550.1238  &  783.0000   & 763.0000  &   10.5396    &    13.0189    &   3.6836   &            & 0.5647    \\
\hline
$a_\sigma$ & $b_\sigma$ & $c_\sigma$ &  $d_\sigma$  &   $a_\omega$  & $b_\omega$ & $c_\omega$ & $d_\omega$\\
  1.3881   &   1.0943   &   1.7057   &    0.4421    &     1.3892    &   0.9240   &    1.4620  &    0.5647 \\
\hline\hline
\end{tabular}
\end{table*}
%-------------------------------------------------------------------

%---------------------------------------------------------------------------
\section{Summary}
\label{sec:summary}
%---------------------------------------------------------------------------

%-------------------------------------------------------------------
\begin{table*}[tb]
\centering
\caption{Alternative parametrization of the density dependence of the
  couplings in the isoscalar channels for the indicated values of
  $K_{\text{sat}}$ (MeV) and/or  $Q_{\text{sat}}$ (MeV). The values of
  $g_{\sigma N}$ and $g_{\omega N}$ are the same as in the DD-ME2 parametrization,
  see Table~\ref{tab:DDME2}.
}
\setlength{\tabcolsep}{10pt}
\label{tab:Varying_KQ}
\centering
\begin{tabular}{cccccccccc}
\hline\hline
$K_{\text{sat}}$&$Q_{\text{sat}}$&$a_\sigma$&$b_\sigma$&$c_\sigma$&$d_\sigma$&$a_\omega$&$b_\omega$&$c_\omega$&$d_\omega$\\
\hline
200 & 480  & 1.4851 & 1.1012 & 1.8753 & 0.4216 & 1.4843 & 0.8786 & 1.5293 & 0.4669 \\
220 & 480  & 1.4469 & 1.1074 & 1.8214 & 0.4278 & 1.4469 & 0.9013 & 1.5110 & 0.4697 \\
240 & 480  & 1.4088 & 1.1015 & 1.7498 & 0.4365 & 1.4096 & 0.9165 & 1.4803 & 0.4745 \\
260 & 480  & 1.3707 & 1.0802 & 1.6573 & 0.4485 & 1.3722 & 0.9212 & 1.4334 & 0.4822 \\
280 & 480  & 1.3328 & 1.0401 & 1.5413 & 0.4650 & 1.3348 & 0.9117 & 1.3670 & 0.4938 \\
300 & 480  & 1.2953 & 0.9756 & 1.3970 & 0.4885 & 1.2976 & 0.8815 & 1.2740 & 0.5115 \\
\\
250 & -600 & 1.3501 & 0.1798 & 0.3299 & 1.0052 & 1.3788 & 0.1467 & 0.2905 & 1.0711 \\
250 & -300 & 1.3406 & 0.3380 & 0.5619 & 0.7702 & 1.3611 & 0.2813 & 0.4915 & 0.8235 \\
250 & 0    & 1.3477 & 0.5546 & 0.8807 & 0.6152 & 1.3612 & 0.4655 & 0.7647 & 0.6602 \\
250 & 300  & 1.3690 & 0.8555 & 1.3353 & 0.4996 & 1.3752 & 0.7205 & 1.1493 & 0.5385 \\
250 & 600  & 1.4077 & 1.2841 & 2.0136 & 0.4069 & 1.4049 & 1.0809 & 1.7136 & 0.4410 \\
250 & 900  & 1.4730 & 1.9201 & 3.0965 & 0.3281 & 1.4571 & 1.6107 & 2.5947 & 0.3584 \\
\\
220 & 0    & 1.3993 & 0.6123 & 1.0183 & 0.5721 & 1.4140 & 0.4990 & 0.8630 & 0.6215 \\
220 & 300  & 1.4244 & 0.8934 & 1.4672 & 0.4766 & 1.4306 & 0.7284 & 1.2282 & 0.5210 \\
220 & 600  & 1.4657 & 1.2752 & 2.1075 & 0.3977 & 1.4610 & 1.0361 & 1.7367 & 0.4381 \\
250 & 900  & 1.5312 & 1.8144 & 3.0791 & 0.3290 & 1.5106 & 1.4654 & 2.4873 & 0.3661 \\
\\
280 & 0    & 1.2987 & 0.4670 & 0.7119 & 0.6843 & 1.3107 & 0.4051 & 0.6354 & 0.7243 \\
280 & 300  & 1.3147 & 0.7809 & 1.1600 & 0.5361 & 1.3206 & 0.6825 & 1.0316 & 0.5684 \\
280 & 600  & 1.3494 & 1.2550 & 1.8662 & 0.4226 & 1.3485 & 1.1020 & 1.6520 & 0.4492 \\
280 & 900  & 1.4143 & 2.0165 & 3.0866 & 0.3286 & 1.4038 & 1.7774 & 2.7166 & 0.3503 \\
\hline\hline
\end{tabular}
\end{table*}
%-------------------------------------------------------------------

Using the EoS for hadronic matter satisfying the latest constraints
from both terrestrial nuclear experiments and astrophysical
observations at saturation, as well as $\chi$EFT of low-density
neutron matter, we found that the gross properties of compact stars
are very sensitive to the higher-order empirical parameters of nuclear
matter around the saturation density, specifically the isoscalar
skewness $Q_{\text{sat}}$ and isovector slope $L_{\text{sym}}$.
These are not well constrained from the experimental side, while $L_{\text{sym}}$
is constrained somewhat by $\chi$EFT.

We observe that the $Q_{\text{sat}}$ is the dominant parameter
controlling both the maximum mass and the radius of a compact star.
This is due to the fact that, on the one hand, the isovector
characteristics $E_{\text{sym}}$ and $L_{\text{sym}}$ in
Eq.~\eqref{eq:Taylor_expansion} weakly influence the maximum
mass, on the other hand, the strong restriction on the allowed
values of $L_{\text{sym}}$ coming from $\chi$EFT does not allow for
noticeable variations in the radius.

Another important point is that the upper limit on $Q_{\text{sat}}$ is
essentially dependent on the assumed particle composition of stellar
matter. Our exploration of the parameter space shows that hyperonic
stars more massive than 2$M_{\odot}$ would require
$Q_{\text{sat}} \gtrsim 200$~MeV, leading to a radius
$R_{M_{1.4}} \gtrsim 12.8$~km. Including in the composition, in
addition, the $\Delta$ resonances reduce the radius of a canonical mass
star by about 1~km for a reasonably attractive $\Delta$-potential,
in agreement with previous findings~\cite{Drago2014a,Lijj2018b,Lijj2019}.

\section*{Acknowledgements}
J.~L.\ acknowledges the support by the Alexander von Humboldt
foundation. A.~S.\ acknowledges the support by the DFG (Grant No.
SE 1836/4-1). Partial support was provided by the European COST
Action ``PHAROS'' (CA16214) and the State of Hesse LOEWE-Program in
HIC for FAIR.

\appendix*
\section{Meson-nucleon coupling constants}

The parameters of the DD-ME2 effective interaction are shown in
Table~\ref{tab:DDME2}. As we already discussed in the text, this model
is well constrained with respect to the characteristics
$\rho_{\text{sat}}$, $E_{\text{sat}}$, and $M^\ast_D$. In
Table~\ref{tab:Varying_KQ} we further present a set of alternative
parametrizations that preserve these values of $\rho_{\text{sat}}$,
$E_{\text{sat}}$, and $M^\ast_D$, but produce different values of
$K_{\text{sat}}$ and/or $Q_{\text{sat}}$. Notice that to this end one
needs to modify only the parameters in functions $f_{\sigma N}$ and
$f_{\omega N}$ [see Eq.~\eqref{eq:isoscalar_coupling}] that control
the density dependence of the couplings in the isoscalar sector.
As the density dependence of the couplings in the isovector sector
is parametrized by an exponential form given by
Eq.~\eqref{eq:isovector_coupling}, the modification for isovector
sector is rather simple: one first determines $g_{\rho N}$ by the
preassigned value of $E_{\text{sym}}$ and then fixes $a_{\rho}$ by
the desired value of $L_{\text{sym}}$.

\bibliographystyle{apsrev}
\bibliography{Comstars_ref}	

\end{document}